\documentclass[12pt]{article}
\usepackage{amsmath}
\usepackage{amssymb}
\usepackage{color}
\usepackage{epsfig}
\usepackage{amsfonts}
\usepackage{xparse}

\makeatletter \@addtoreset{equation}{section} \makeatother
\makeatletter \@addtoreset{figure}{section} \makeatother

\def\IC{\mathbb{C}}
\def\IP{\mathbb{P}}
\def\IZ{\mathbb{Z}}

\def\CH{{\cal H}}

\def\CO{{\cal O}}

\def\CT{{\cal T}}

\newcommand{\beq}{\begin{equation}}
\newcommand{\eeq}{\end{equation}}
\newcommand{\bea}{\begin{eqnarray}}
\newcommand{\eea}{\end{eqnarray}}
\newcommand{\ba}{\begin{array}}
\newcommand{\nn}{\nonumber}
\newcommand{\bi}{\begin{itemize}}
\newcommand{\ei}{\end{itemize}}

\NewDocumentCommand{\ce}{s O{} m}{%
  \IfBooleanTF{#1} 
    {\left\lceil#3\right\rceil} 
    {#2\lceil#3#2\rceil} 
}

\NewDocumentCommand{\fl}{s O{} m}{%
  \IfBooleanTF{#1} 
    {\left\lfloor#3\right\rfloor} 
    {#2\lfloor#3#2\rfloor} 
}

\def\ba{{\bar a}}

\def\bi{{\bar i}}

\begin{document}
\begin{titlepage}
\vfill
\begin{flushright}
{\tt\normalsize KIAS-P12037}\\

\end{flushright}
\vfill
\begin{center}
{\Large\bf BPS States, Refined Indices,  \\\vskip 2mm
and  Quiver Invariants }

\vskip 1cm
Seung-Joo Lee,\footnote{\tt s.lee@kias.re.kr}
Zhao-Long Wang,\footnote{\tt zlwang@kias.re.kr}
and Piljin Yi\footnote{\tt piljin@kias.re.kr}

\vskip 5mm
{\it School of Physics, Korea Institute
for Advanced Study, Seoul 130-722, Korea}

\end{center}
\vfill

\begin{abstract}
\noindent
For $D=4$ BPS state construction, counting, and wall-crossing thereof,
quiver quantum mechanics offers two alternative approaches, the Coulomb
phase and the Higgs phase, which sometimes produce inequivalent counting.
The authors have proposed, in arXiv:1205.6511, two conjectures on the
precise relationship between the two, with some supporting evidences. Higgs
phase ground states are naturally divided into the Intrinsic Higgs sector,
which is insensitive to wall-crossings and thus an invariant of quiver,
plus a pulled-back ambient cohomology, conjectured to be an one-to-one
image of Coulomb phase ground states. In this note, we show that
these conjectures hold for all cyclic quivers with Abelian nodes,
and further explore angular momentum and $R$-charge content of individual
states. Along the way, we clarify how the protected spin character
of BPS states should be computed in the Higgs phase, and further determine
the entire Hodge structure of the Higgs phase cohomology. This shows
that, while the Coulomb phase states are classified by angular momentum,
the Intrinsic Higgs states are classified by $R$-symmetry.
\end{abstract}

\vfill
\end{titlepage}

\tableofcontents\newpage
\renewcommand{\thefootnote}{\#\arabic{footnote}}
\setcounter{footnote}{0}

\section{Introduction}

In the study of four-dimensional $N=2$ supersymmetric theories, one is naturally
led to supersymmetric quiver quantum mechanics \cite{Denef:2002ru},
normalizable ground states of which are images of quantum BPS
particle states~\cite{Prasad:1975kr} of the underlying theory in
four dimensions. In the language of quiver quantum mechanics,
wall-crossings \cite{Seiberg:1994rs,Ferrari:1996sv} occur
when Fayet-Iliopoulos (FI) constants of the quiver change signs,
upon which the number of normalizable ground states changes discontinuously.
The Coulomb phase realizes a more physical and intuitive picture of
this process via multi-centered nature of typical BPS states;
some BPS states disappear simply because they are bound states of more than one
charge centers and these states become non-normalizable
at such crossings
\cite{Lee:1998nv,Bak:1999da,Gauntlett:1999vc,Denef:2000nb,Ritz:2000xa,Argyres:2001pv}.

Such discontinuities are also mirrored by the Higgs phase, but in a less
intuitive manner. Depending on signs of FI constants, one ends up in
different branches, labeled by $k$ in this note, with different vacuum
manifolds, say $M_k$. Each of these branches has
its own supersymmetric ground states represented by the cohomology ring, $H(M_k)$
\cite{Denef:2002ru}. The relevant index in its simplest form is
\begin{equation}
(-1)^{-d_k}\chi(M_k)=\sum_l (-1)^{l-d_k}\;{\rm dim}\;H^l(M_k)\ ,
\end{equation}
with $d_k = {\rm dim}_\IC M_k$,
whose refined form as well as the Coulomb phase counterpart will
be introduced in section 2 and explored thereafter. Although
such indices are supposed to be invariant under small continuous
deformations, these actually jump discontinuously across
marginal stability walls, defined by vanishing of
certain FI constants, simply because the respective vacuum
manifolds $M_k$ are topologically distinct in different branches.

Since
quiver theories in question are quantum mechanics, what we call
phases here do not have the same meaning as in field theories. In
the latter, with more than two spacetime dimensions, phases imply
super-selection sectors, as small fluctuations cannot change vacuum
expectation values (vev). In quantum mechanics, there is no such
notion as vev, so Higgs and Coulomb phases are just different
manifestations or approximations of the same ground state sector physics.
Thus, in the absence of some further subtlety, it is natural to
expect that Coulomb phase states are mirrored by Higgs phase states
and, in one large class of examples, this expectation was confirmed by
extensive comparisons \cite{Denef:2002ru} and more recently
by a general and analytical proof \cite{Sen:2011aa}.

However,
a curious fact is that, when the quiver involves a closed loop,
$\chi(M_k)$ sometimes predicts
far more numerous Higgs phase ground states  than those found in
the corresponding Coulomb phase \cite{Denef:2007vg}. Upon closer
inspections of the quiver dynamics \cite{Kim:2011sc,Lee:2011ph},
this disparity between the two phases is not entirely surprising, as the Coulomb phase
has some subtleties that could in principle compromise the usual
approach; see sections 2 and 7 for more detail. Since
(dis-)appearance of BPS states across the walls is clearly
due to the multi-centered nature, which is characteristic of
the Coulomb phase, these additional states in the Higgs phase are
likely unaffected by wall-crossings. On this expectation, it has been suggested
that such extra Higgs phase states must have something to do with
single-centered black hole solution \cite{Manschot:2011xc}.  Whatever the
right spacetime interpretation might
be, a more immediate question is how should Higgs phase ground states
for general quivers be classified in relation to their Coulomb phase cousins.
That is, how to identify the Higgs phase counterpart of the
Coulomb phase states, and what are distinguishing geometric
and quantum mechanical properties of those extra Higgs phase states
that have no Coulomb phase counterpart.

In Ref.~\cite{Lee:2012sc}, the authors proposed two conjectures,
starting with the observation that Higgs phase vacuum
manifolds, $M$, are always of the form
\begin{equation}
M=X\biggl\vert_{\partial W=0}\quad \hookrightarrow\quad X \ ,
\end{equation}
where $X$ is a D-term-induced K\"ahler manifold. $W$ is the
superpotential, from which F-term constraints $\partial W=0$ are generated.
Geometrically, this means that $M$ is a complete intersection in
the ambient projective variety $X$.
The cohomology of $M$ is naturally split as
\begin{equation}
H(M)=i^*_M(H(X))\oplus \left[H(M)/i^*_M(H(X))\right] ,
\end{equation}
where $i_M: M\hookrightarrow X$ is the embedding map. Labeling
branches by the subscript $k$ as before, the conjectures can be stated as
\begin{itemize}

\item
The states in the pull-back $i^*_{M_k}(H(X_k))$ of the ambient cohomology are in
one-to-one correspondence with the Coulomb phase states. In terms of
the equivariant index $\Omega_{\rm Coulomb}^{(k)}(y) $, we assert that
\begin{equation}\label{EquivConj1}
\Omega_{\rm Coulomb}^{(k)}(y)=(-y)^{-d_k}D_k(-y)\ ,
\end{equation}
where, on the right hand side coming from Higgs phase,  $d_k$ is the
complex dimension of $M_k$ and $D_k(x)$ is the reduced Poincar\'{e} polynomial
defined as
\begin{equation}\label{1}
D_k(x)\equiv \sum_l x^l \,{\rm dim}\left[i^*_{M_k}(H^l(X_k))\right] \ .
\end{equation}

\item
The Intrinsic Higgs states in $H(M_k)/i^*_{M_k}(H(X_k))$ are inherent
to the  quiver quantum mechanics and insensitive to wall-crossing.
In particular, their numerical index, given by
\begin{equation}\label{2}
(-1)^{-d_k}\chi(M_k)-(-1)^{-d_k} D_k(-1)\,,\end{equation}
is independent of $k$, unchanged by wall-crossings.\footnote{Since the
parity $(-1)^{d_k}$ is independent of the branch choice $k$, we will
sometimes drop this overall sign factor, as was done in Ref.~\cite{Lee:2012sc};
restoring the sign to the current form is a trivial exercise.} More
generally, this extends to a refined index of the form
\begin{equation}\label{2'}
(-y)^{-d_k}\chi_{\xi = -y^2}(M_k)- (-y)^{-d_k} D_k(-y)\ ,
\end{equation}
which defines an invariant of the quiver. Here, $\chi_\xi$ is the refined
Euler character, to be introduced in section 6.

\end{itemize}
In Ref.~\cite{Lee:2012sc}, an analytical proof of (\ref{EquivConj1})
was given for cyclic Abelian quivers with three nodes,\footnote{For
three-node examples, Bena et.al. \cite{Bena:2012hf} independently
explored the same phenomena.} while the invariance of (\ref{2}) was
shown analytically for three node cases and at numerical level
for four, five, and six node examples. In this note, we wish to
work towards complete proofs of the conjectures
and explore related issues.

It turns out that, for cyclic Abelian quivers with arbitrary number of nodes,
proofs are relatively easy and compact, so those will be the main content
of this note. Along the way, we also discuss how the so-called
protected spin character \cite{Gaiotto:2010be} of $D=4$ $N=2$
supersymmetric theories, carrying both angular momentum
and $SU(2)_R$ representations of individual BPS state,
should be computed in the Higgs phase. The precise relationship
between the protected spin character
and the Coulomb phase equivariant index was clarified very
recently \cite{Kim:2011sc}, and here, we extend this to the Higgs phase.
Interestingly, we find that the Intrinsic Higgs states, all in singlets of
angular momentum $SU(2)_J$, are classified by R-charges.

The organization of this note is as follows.
In section 2, we start with a broad overview of refined indices
from the underlying four-dimensional theories and discuss how they
can be computed in the quiver realization of BPS states. After
a brief review of how the protected spin character of the former
reduces to the equivariant index of the latter in the Coulomb phase,
we  propose the Higgs phase version. This in part will justify
the  refined form of the first conjecture, which is proved
later in this note. In section 3, we review basics of the Higgs phase
moduli spaces for cyclic Abelian quivers and their cohomologies,
in preparation for subsequent discussions. In section 4,
we take advantage of the relatively simple form of the
reduced Poincar\'e polynomial $D_k(x)$ and show how it is precisely
mapped to the Coulomb phase equivariant index, thereby providing
a proof of the first conjecture.

In section 5, we explore a bit more about
the pulled-back cohomology $i^*_{M_k}(H(X_k))$ and give a
simple geometric realization. Although a fully general
proof of the second conjecture at the refined level will be given
in section 6, this geometric view allows an intuitive
understanding of why the number of states in the Intrinsic Higgs
sector is an invariant. As a byproduct, we also find an elementary
arithmetic counting formula for the degeneracy,
$D_k(-1)=(-1)^{d_k}\Omega_{\rm Coulomb}^{(k)}(1)$.
In section~6, we come back to the  refined index of the Higgs
phase as proposed in section~2. A general computing
algorithm for the refined index is presented and its properties are
illustrated with diverse examples. Thanks to special nature of both
Higgs and Coulomb phase states, these refined indices allow us
to determine the entire Hodge structure of the cohomology, in particular.
As expected on general grounds, we find typically very large degeneracy
of the Intrinsic Higgs sector.
Finally, the conjectured invariance of the Intrinsic Higgs sector
at such refined level is established rigorously. In section 7,
we close with comments and further questions.

\section{Protected Spin Character, Symmetries, and Refined Indices  }

Before we start discussing the conjectures, their proofs, and consequences thereof,
let us briefly recall the definitions of the relevant refined indices. This will
also provide a strong motivation for the two conjectures outlined in the
introduction and, in particular, show how the refined indices are to be
identified across the two phases.

Those states that we wish to count are  particle-like BPS
states of four-dimensional $N=2$ theories, be they Seiberg-Witten,
supergravity, or compactified superstring theories. Since $N=2$
supersymmetry comes with $SU(2)_R\times U(1)_R$ $R$-symmetry group, one might
naturally expect $SU(2)_J\times SU(2)_R\times U(1)_R$ to be the relevant
global symmetry for a given particle-like state,
where $SU(2)_J$ is the spatial rotation. Of these,
however, $U(1)_R$ is ``spontaneously broken" by the phase of the relevant
central charge, so  BPS states would be in
a representation under $SU(2)_J\times SU(2)_R$ as
\begin{equation}
[{\rm1/2\; hyper}]\otimes (J,I) \ ,
\end{equation}
where the first, universal factor consists of a single $SU(2)_J$ doublet and a pair of
singlets. $J$ and $I$ in the second factor denote the spin and the iso-spin
representations, respectively, under $SU(2)_J$ and $SU(2)_R$.

The simplest index for BPS states in $N=2$ theories is the second helicity trace
\begin{equation}
\Omega_2=-\frac12\,{\rm Tr}\; (-1)^{2J_3}(2J_3)^2 \ ,
\end{equation}
where the trace is over the one-particle Hilbert space of
a given charge. This simplifies to
\begin{equation}
\Omega_2={\rm tr}\; (-1)^{2J_3} \ ,
\end{equation}
when we factor out the universal ``center of mass" degrees of
freedom. In practice, this is reflected on the lowercase ``tr"
which ignores the 1/2 hyper part.  As is clear from this, a
half-hyper multiplet would contribute $+1$ and a vector multiplet, $-2$, and so on.
More generally, this index can be elevated to the protected spin
character \cite{Gaiotto:2010be}, as
\begin{equation}
\Omega(y)={\rm tr} \;(-1)^{2J_3}y^{2J_3+2I_3} \ ,
\end{equation}
which ultimately is the object that we wish to compute and compare
in the Higgs and the Coulomb phases.

The story of how this protected spin character reduces to
refined indices of quiver quantum mechanics can be surprisingly
subtle.
For the Coulomb phase, where the bi-fundamental chiral fields are
integrated out, the remaining degrees of freedom are the position
three-vectors $\vec x_i$ of the $n+1$ charge centers and the
fermionic partners thereof, four real for each $i$.
In this phase, it has been shown that there is actually an $SO(4)$
global symmetry \cite{Lee:2011ph,Kim:2011sc}. The three-vectors and the fermionic partners are
in $(\bold 3,\bold 1)$ and $(\bold 2,\bold 2)$, respectively, under $SO(4)= SU(2)_J\times
SU(2)_R$. As such,
the supercharges are also in $(\bold 2,\bold 2)$ and the protected spin character
descends to the $\mathbb{R}^{3(n+1)}$ Coulomb phase quantum mechanics, verbatim,
\begin{equation}
\Omega_{\rm Coulomb}(y)={\rm tr} \;(-1)^{2J_3}y^{2J_3+2I_3} \ .
\end{equation}
The manifestation of $SU(2)_J\times SU(2)_R$ symmetry in the Coulomb
phase is related to the fact that this picture can be directly derived
from low energy dynamics of quantum solitons (or black holes) in four-dimensional
$N=2$ theories \cite{Lee:2011ph}.

A more subtle step, which had caused some confusions in literature
in the past, occurs when one tries to formulate this index problem
in terms of the classical moduli space determined by Denef's  formulae,
\begin{equation}
\zeta_i =\sum_{j\neq i}\frac{\langle \gamma_{i},\gamma_j \rangle}{|\vec x_i-\vec x_j|} \ ,
\end{equation}
which carve out a $2n$-dimensional submanifold in the ``relative
position space''  $\mathbb{R}^{3n}$ spanned by $\vec x_i-\vec x_j$'s.
As will be commented upon later, $\zeta_i$'s are the FI constants of
the quiver theory, while the numerators on the right hand side
are the linking numbers which count bi-fundamental fields,
or equivalently the Schwinger products of the charge pairs.
Although it appears natural to reduce dynamics onto such a classical
moduli space, no such consistent truncation exists in the quiver
theory in general. The main problem is that quantum gaps along
classical moduli space are always equal to the gaps
along the classically massive directions \cite{Kim:2011sc}. This
is one main qualitative difference between the Coulomb phase
dynamics  and  the Higgs phase dynamics, as no such subtlety shows
in the Higgs phase.

Instead, one may deform the dynamics by hand to make the
classical gaps along ``radial" directions, $|\vec x_i-\vec x_j|$,
to be much larger than the quantum gaps along the classically
flat ``angular" directions.
This deformation preserves one out of four supercharges, say
${\cal Q}$, and also preserves  the diagonal combination of
$SU(2)_J\times SU(2)_R$. Calling the latter $SU(2)_{\cal J}$ with ${\cal J}=J+I$, it has been
shown \cite{Kim:2011sc} that the protected spin character thus
reduces to the refined index of the form\footnote{Strictly speaking, the overall sign in
this formula is valid for Abelian quivers only. Its
generalization to cases with more than one identical
particles is rather involved. We refer readers
to \cite{Kim:2011sc,Manschot:2010qz,Manschot:2011xc}.}
\begin{equation}
\Omega_{\rm Coulomb}(y)=(-1)^{-n+\sum\limits_{i>j}\langle
 \gamma_i,\gamma_j\rangle}{\rm tr} \;(-1)^{F}y^{2{\cal J}_3} \ .
\end{equation}
Here, the chirality operator $(-1)^F$, which anticommutes with
the surviving supercharge ${\cal Q}$, can be thought of as
$(-1)^{2I_3}$; although $SU(2)_R$ is no longer a symmetry,
its discrete remnant $(-1)^{2I_3}$ still is. Furthermore, this chirality
operator  coincides with the canonical choice in mathematics literature,
with which standard index theorems are stated. The
resulting index is what has been identified as the equivariant
index of the Coulomb phase, $\Omega_{\rm Coulomb}(y)$.
See Refs.~\cite{Gauntlett:1999vc,Stern:2000ie,Denef:2002ru,
Denef:2007vg,deBoer:2008zn,Manschot:2010qz,Lee:2011ph,
Manschot:2011xc, Kim:2011sc} for physical derivations
and computations, in one limit or another.

An important consistency check of the above refined index
is that $y^{2{\cal J}_3}$ should  commute with the surviving supercharge.
In other words, there should be at least one supercharge, ${\cal Q}$,
such that ${\cal Q}^2$ can act as the Hamiltonian and
\begin{equation}
\{(-1)^Fy^{2{\cal J}_3}, {\cal Q}\}=0\ .
\end{equation}
The latter is possible only if  $[{\cal J}_3,{\cal Q}]=0$, since $y$ is arbitrary.
Otherwise, the above refined index itself makes no sense. In the Coulomb phase
dynamics, this comes about because the supercharges in $(\bold 2,\bold 2)$
under $SU(2)_J\times SU(2)_R$ decompose into $\bold 3+\bold 1$ under
the diagonal $SU(2)_{\cal J}$, and because the deformation preserves precisely
the singlet supercharge, ${\cal Q}$, from which  $[{\cal J}_3, {\cal Q}]=0$
follows \cite{Kim:2011sc}.

Before moving on to the Higgs phase, let us note that all BPS states
constructed to date from the multi-center picture are $SU(2)_R$
singlets.\footnote{For an extensive list of examples from explicit
field theory constructions,  see Ref.~\cite{Weinberg:2006rq}.} Although
the refined Coulomb phase index is defined with the grading $2{\cal J}_3$,
this observation leads us to believe that $y^{2{\cal J}_3}$ is effectively
the same as $y^{2J_3}$ in the supersymmetric ground state sector. It may
as well be that $\Omega_{\rm Coulomb}(y)$ simply keeps track of angular
momentum content of Coulomb phase BPS states \cite{Gaiotto:2010be}.
Nevertheless, it is important to remember that the refined index
should be defined with the $y^{2{\cal J}_3}$ insertion;
otherwise, the expression would not reduce to the ground state sector
and thus will not define an index.

For the Higgs phase expression of the protected spin character,
first recall that, there, the dynamics is a supersymmetric
nonlinear sigma model onto a complete intersection $M\hookrightarrow X$
inside a projective variety $X$. In contrast to the Coulomb phase,
reduction of dynamics down to the classical moduli space
is straightforward as long as FI constants are of large enough absolute
values.  $M$ inherits the K\"{a}hlerian
properties from the quiver data and is automatically equipped with
$SU(2)_{\rm Lefschetz}$ symmetry acting on wavefunctions.
The three generators of $SU(2)_{\rm Lefschetz}$ can be represented as~\cite{GH}
\begin{equation}
L_3=(l-d)/2\ , \quad L_+=K\wedge  \ , \quad L_-=K \lrcorner\,\,\,,
\end{equation}
when acting on the cohomology $H(M)=\bigoplus_l H^l(M)$. Here,
$d={\rm dim}_\IC M$ is the complex dimension of the vacuum manifold,
as before, and $K$ is the K\"ahler two-form.

In the past, the grading by $2L_3$ in the Higgs phase index,
manifest in the Poincar\'e polynomial of $M$ for example, has
been successfully matched with the grading by $2{\cal J}_3$
in $\Omega_{\rm Coulomb}(y)$ \cite{Denef:2002ru,Manschot:2010qz,Sen:2011aa}.
On the other hand, since $\Omega_{\rm Coulomb}$'s that
have been so far used in such comparisons   contain
only $SU(2)_R$ singlet states, this leaves an ambiguity
of whether $L={\cal J}$ or $L=J$. What fixes the matter
once and for all is how supercharges transform under
$SU(2)_{\rm Lefschetz}$; as we will see below, from Eq.~(\ref{LI}),
it is clear that $y^{2L_3}$ cannot commute with any of the four
supercharges. This shows that $SU(2)_{\rm Lefschetz}=SU(2)_J$
and $L=J$; We must find an analog of ${\cal J}_3$, or equivalently
an analog of $I_3$, to compute the protected spin character
in the Higgs phase.

Recall that all quiver theories are equipped with a $U(1)_R'$ symmetry,
to be distinguished from $U(1)_R$ of the underlying four-dimensional $N=2$
theories. This symmetry is already evident in and inherited from
$D=4$ version of the quiver theories, so $U(1)_R'$ clearly
commutes with the spatial rotation group $SU(2)_J$.
Upon the Hodge decomposition of differential forms,
and in particular, of the ground state sector,
\beq
H(M)=\bigoplus\limits_l H^l(M)=\bigoplus\limits_{p,q} H^{p,q}(M) \ ,
\eeq
the obvious
$U(1)_R'$ charge assignment is
\begin{equation}
{\cal I} = (p-q)/2 \ ,
\end{equation}
on $H^{p,q}$, from which it immediately follows $[{\cal I}, L_{1,2,3}]=0$.
With $SU(2)_{\rm Lefschtez}=SU(2)_J$, $U(1)_R'$ generated by ${\cal I}$ must then be
a remnant of $SU(2)_R$ of the underlying four-dimensional $N=2$ supersymmetric
theories. This leads us to propose that
the protected spin character is computed in the Higgs phase as
\begin{eqnarray}
\Omega_{\rm Higgs}(y) &=& {\rm tr}\;(-1)^{2L_3}y^{2L_3+2{\cal I}} \nonumber\\
&=& {\rm tr} \;(-1)^{\,l-d}y^{\,l-d+p-q}\nonumber \\
&=& {\rm tr} \;(-1)^{p+q-d}y^{2p-d} \ ,
\end{eqnarray}
where for the last equality we used $l=p+q$.

Again, an important consistency  check here is the existence of a
supercharge, ${\cal Q}$ (in the Higgs phase, together with
its Hermitian conjugate ${\cal Q}^\dagger$), which commutes with $L_3+{\cal I}$.
For this, let us divide the complex fermions into holomorphic $\phi^\alpha$,
anti-holomorphic $\bar \chi^{\bar\beta}$, and their conjugates
$\phi_{\alpha}^\dagger$ and  $\bar\chi_{\bar\beta}^\dagger$, respectively. The
canonical supercharge is schematically,
\begin{equation}
{ Q}\;\;\sim \;\;\phi^\alpha\partial_\alpha +\bar\chi^{\bar\beta}\partial_{\bar\beta} \ .
\end{equation}
Treating $\phi$ and $\bar\chi$ as the creation operators among fermions
and adopting the usual differential form representation of states, we may
identify
\begin{equation}
{Q}=\partial+\bar \partial\ ,
\end{equation}
where $\partial$ ($\bar\partial$) is the (anti-)holomorphic exterior
derivative. In this language, we have
\begin{equation}\label{LI}
L_3= \frac12\left(\phi\phi^\dagger+\chi\chi^\dagger-d\right)\ , \qquad
{\cal I}= \frac12\left(\phi\phi^\dagger-\chi\chi^\dagger\right)\ ,
\end{equation}
which shows that $L_3 +{\cal I}= \phi\phi^\dagger-d/2$ commutes with
both $\bar\partial$ and its adjoint $\bar\partial^\dagger$.
Therefore, we have a complex supercharge pair ${\cal Q}=\bar\partial$ and
 ${\cal Q}^\dagger=\bar\partial^\dagger$,
with respect to which $\Omega_{\rm Higgs}(y)$ is a
refined index.

Finally,  recall how our conjectures naturally split the cohomology
of $M$ in the Higgs phase into the Intrinsic Higgs sector and the
pull-back $i^*_M(H(X))$ of the ambient cohomology. In all of our
examples, the ambient space $X$ has a simple cohomology structure
\beq
H(X)=\bigoplus_{p} H^{p,p}(X) \ ,
\eeq
corresponding to the statement that $i^*_M(H(X))$ contains
only $SU(2)_R$ singlets (or $U(1)_R'$ neutral).
Then, the pulled-back part of the Higgs  phase index can be expressed as
\begin{eqnarray}
\Omega_{\rm Higgs}(y)\biggr\vert_{i^*_M(H(X))}
&=& {\rm tr}_{i^*_M(H(X))}\;(-1)^{2L_3}y^{2L_3}\nn \\
&=& (-y)^{-d}\; {\rm tr}_{i^*_M(H(X))} \;(-y)^l \nn \\
&=& (-y)^{-d}D(-y) \ .
\end{eqnarray}
Here, in the last step, the reduced Poincar\'e
polynomial $D(x)$, defined in Eq.~\eqref{1}, appears,
motivating the refined version of the first conjecture as stated in
Eq.~\eqref{EquivConj1}. The Intrinsic Higgs sector, on the other hand,
is constrained to the middle cohomology by the Lefschetz hyperplane
theorem \cite{Lee:2012sc,GH} and the refined index restricted there
becomes
\begin{eqnarray}
\Omega_{\rm Higgs}(y)\biggr\vert_{\rm Intrinsic}
&=& {\rm tr}_{H(M)/ i^*_M(H(X))}\;y^{2{\cal I}} \nonumber\\
&=& {\rm tr}_{H(M)/ i^*_M(H(X))} \;y^{2p-d} \ .
\end{eqnarray}
The Intrinsic Higgs
sector belongs to the middle cohomology $p+q=d$, {i.e.}, states
in this sector are all angular momentum singlets.
This restriction to the middle cohomology is valid for irreducible
cases where $M\hookrightarrow X $ does not factorize as
$(M'\hookrightarrow X')\times Y $.  Generalization to reducible cases
is straightforward.

\section{Higgs Phase Cohomologies: Cyclic $(n+1)$-Gon Quivers}

\begin{figure}[!h]
\centering
\includegraphics[width=9cm]{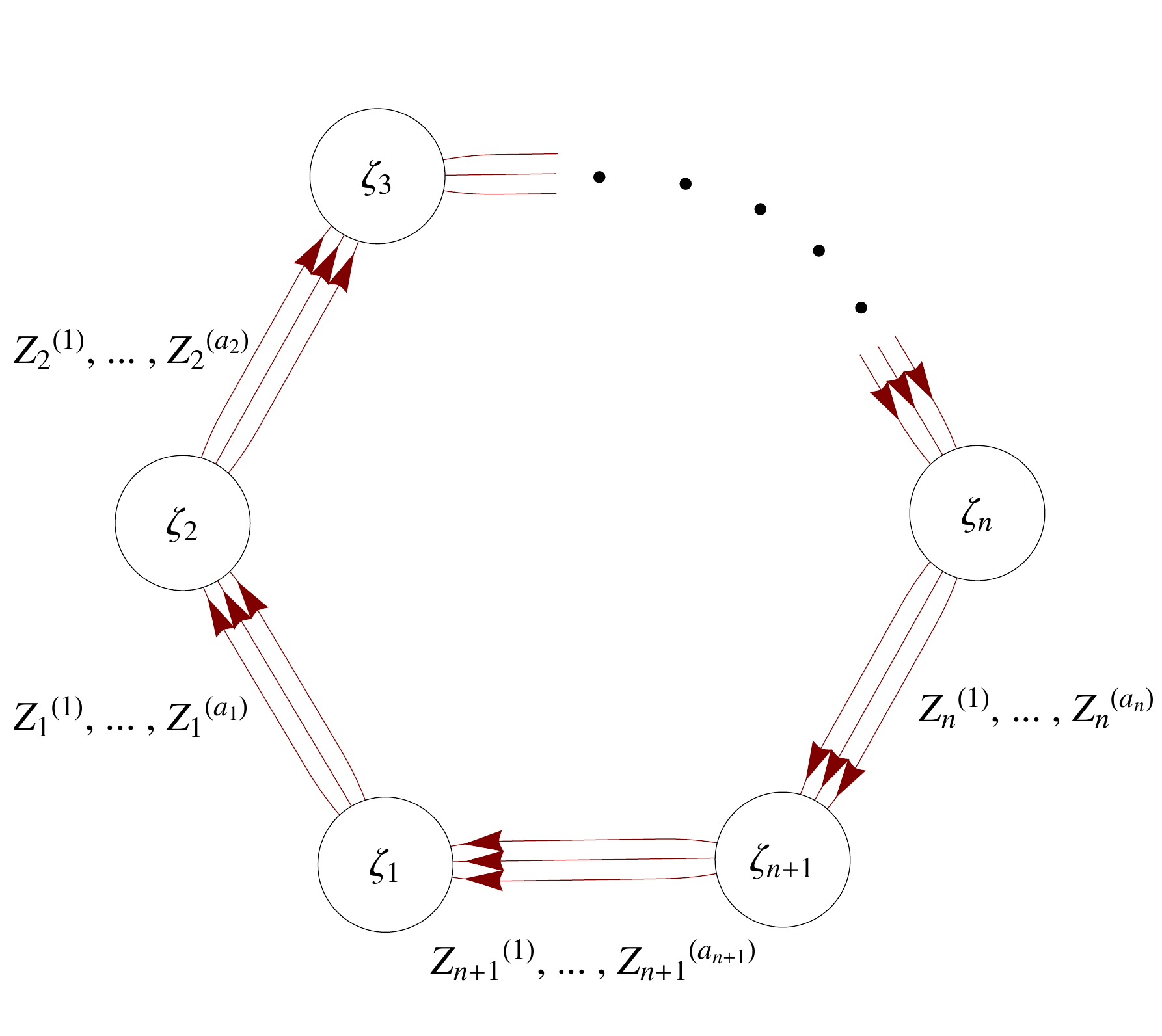}
\parbox{6in}{\caption{\small The figure, borrowed from Ref.~\cite{Lee:2012sc}, shows
a cyclic Abelian quiver with $n+1$ nodes. Associated with each node is a FI constant $\zeta_i$
and a $U(1)$ gauge field. Arrows between $i$-th and $(i+1)$-th nodes represent
$a_i$ chiral multiplets of charge $(-1,1)$, say $Z_i = (Z_i^{(1)}, \cdots, Z_i^{(a_i)})$.}\label{fig:quiver}}
\end{figure}

Let us start with a cyclic $(n+1)$-gon quiver and denote by $Z_{i}$ the
bi-fundamental fields connecting $i$-th and $(i+1)$-th nodes, for $i=1, \dots, n$.
The last, connecting $(n+1)$-th node to the first, is called ${Z}_{n+1}$.
For each $i$, there are $a_i$ arrows from the $i$-th node to the $(i+1)$-th node,
encoding the fact that $Z_i$ is an $a_i$-dimensional complex vector,
\begin{equation}
Z_i = (Z_i^{(1)}, \cdots, Z_i^{(a_i)}) \ ,
\end{equation}
which has charge $(-1,1)$ with respect to the two $U(1)$'s.
See Figure~\ref{fig:quiver}.
Because the quiver has a loop, with  linking numbers of the same sign,
we should expect a generic superpotential of the type,
\begin{equation}
W({Z}_1,{Z}_2,\cdots,{Z}_{n+1})=\sum_{\beta_1=1}^{a_1}\cdots
\sum_{\beta_{n+1}=1}^{a_{n+1}} c_{\beta_1\beta_2\cdots \beta_{n+1}}
{Z}^{(\beta_1)}_1{Z}^{(\beta_2)}_2\cdots{Z}^{(\beta_{n+1})}_{n+1}\,,
\end{equation}
whose F-term vacuum conditions are
\begin{equation}
\partial_{{Z}^{(\beta_i)}_i}W=0~~~\,(\beta_i=1,2,3,\cdots,a_i)\,,
\end{equation}
for $i=1,\cdots,n+1$. As argued in Refs. \cite{Denef:2007vg,Lee:2012sc},
solutions to $\partial W=0$ split into branches, where one of the
complex vectors ${Z}_{i}$'s is identically zero.

The choice of branch is dictated by the D-term constraints,
on the other hand. With $U(1)^{n+1}$ gauge groups, of which
the overall sum decouples, we have $n$-independent
D-term conditions
\begin{eqnarray}
|Z_{n+1}|^2-|Z_1|^2&=&\zeta_1\,,\\ \nonumber
|Z_1|^2-|Z_2|^2&=&\zeta_2\,,\\ \nonumber
|Z_2|^2-|Z_3|^2&=&\zeta_3\,,\\ \nonumber
&\vdots& \\ \nonumber
|Z_n|^2-|Z_{n+1}|^2&=&\zeta_{n+1}\,,\\ \nonumber
\end{eqnarray}
with
\begin{eqnarray}
\zeta_1+\cdots+\zeta_{n+1}=0\,.
\end{eqnarray}
The $k$-th branch is realized when
\begin{equation}
\sum_{i= I}^{k} \zeta_{i}> 0\ ,\quad \sum_{i= k+1}^{ J}\zeta_i<0\ ,
\end{equation}
for consecutive and mutually exclusive sets of $I$'s and $J$'s,
where the cyclic nature of the indices are understood. In the
$k$-th branch, ${Z}_{k}$ is a zero vector and the remaining D-term conditions are solved entirely by
$$
X_k={\mathbb{CP}}^{a_1-1}\times\cdots\times {\mathbb{CP}}^{a_{k-1}-1}
\times {\mathbb{CP}}^{a_{k+1}-1}\times\cdots\times {\mathbb{CP}}^{a_{n+1}-1}\ ,
$$
where individual sizes of $\mathbb{CP}^{a_i-1}$ do not matter, as long
as they are all nonzero.
At the boundaries of a  branch, one or more of $\mathbb{CP}$'s
get squashed to zero size, and wall-crossing may occur.

Going back to the F-term constraints, we learn that $M_k$
has the form of a complete intersection,
\begin{equation}
M_k=X_k\biggl\vert_{\partial_{{Z}_k}W=0} \quad\hookrightarrow\quad X_k\ ,
\end{equation}
given by zero-locus of $a_k$ F-terms, $\partial_{{Z}_k}W=0 $;
all other F-term conditions are trivially met by  $Z_k=0$.
For $M_k$ in question, we have the complex dimension
\begin{equation}
d_k=\sum_{i\neq k}a_i-a_k -n=\sum_{i=1}^{n+1} a_i -2a_k-n \ .
\end{equation}
The Higgs phase ground states are represented by the cohomology
$H(M_k)$, and the Poincar\'{e} polynomial encodes their counting,
\begin{equation}
P[M_k](x)=\sum_{l=0}^{2d_k} b_l(M_k) \;x^l
\end{equation}
with Betti numbers, $b_l(M_k)$. Recalling
$SU(2)_{\rm Lefschetz}=SU(2)_J$, we see that
$b_l$ counts the Higgs phase ground states of helicity $(l-d_k)/2$.

Our conjectures split the cohomology $H(M_k)$ into two parts,
one of which comes from the pull-back of $H(X_k)$. As used extensively in Ref.~\cite{Lee:2012sc},
the Lefschetz  hyperplane theorem implies that the pull-back of $H(X_k)$ is
isomorphic to $H(M_k)$ for the lower-half cohomologies of $M_k$,
i.e. up
to $H^{d_k-1}$. Also, the map is injective for $H^{d_k}$.
Combined with the Poincar\'{e} duality on $M_k$, this tells us that
\begin{equation}
H^l(M_k)\simeq H^{2d_k-l}(M_k)\simeq H^l(X_k)  \ , \qquad l < d_k \ ,
\end{equation}
or equivalently,
\begin{equation}
H(M_k)=i^*_{M_k}(H(X_k))\oplus [H^{d_k}(M_k)/i^*_{M_k}(H^{d_k}(X_k))]\,.
\end{equation}
Thus,  we  learn that
the Intrinsic Higgs states all belong to the middle cohomology of $M_k$.\footnote{
For more general quivers, we may have reducible cases, where $M=M'\times Y$ and
$M'\hookrightarrow X'$ with projective varieties $X'$ and $Y$.
If reducible, we simply have a product structure $H(M)=H(M')\otimes H(Y)$,
and much of our discussion from now on should apply to $H(M')$.}
Upon the Hodge decomposition, the pull-back occupies the vertical
center line of the Hodge diamond
\begin{equation}
i^*_{M_k}(H(X_k))\;\; \subset\;\; \bigoplus_p H^{p,p}(M_k) \ .
\end{equation}
All other entries, $H^{p,q}(M_k)$ with $p\neq q$ and $p+q\neq d_k$,
are null. This generic feature will allow us to determine the
individual $H^{p,q}(M_k)$ in section 6.

Full information about $i^*_{M_k}(H(X_k))$ can be found from the
well-known Poincar\'{e} polynomial of $X_k$,
\begin{equation}\label{poincare-poly}
P[{X_k}](x)
= \frac{\prod_{i\neq k}(1-x^{2a_i})}{(1-x^2)^n}=\sum b_{2l}(X_{k})\cdot x^{2l} \ .
\end{equation}
We truncate this up to the order $d_k$, and  complete
it by inverting the lower half to the upper half, which computes
the reduced Poincar\'{e} polynomial for $M_k\hookrightarrow X_k$,
\begin{eqnarray}\label{PXM}
D_k(x) &\equiv& \sum_l x^l \,{\rm dim}\left[i^*_{M_k}(H^l(X_k))\right] \nonumber \\
&=&b_{d_k}(X_{k})\cdot x^{d_k}+\sum_{0\le 2l<d_k}  b_{2l}(X_{k})\cdot(x^{2l}+x^{2d_k-2l})\ ,
\end{eqnarray}
where $b_{d_k}(X_{k})$ is nonvanishing only when $d_k$ is even.
Since $i^*_{M_k}(H(X_k))$ belongs to $\bigoplus_p H^{p,p}(M_k)$,
this reduced polynomial determines $i^*_{M_k}(H(X_k))$ entirely.

\section{Coulomb Inside Higgs: the First Conjecture}

In this section, we  prove the first conjecture
\begin{equation}
\Omega_{\rm Coulomb}^{(k)}(y)=(-y)^{-d_k}D_k(-y)\,,
\end{equation}
for all cyclic Abelian quivers.  What facilitates  the proof greatly
is the observation that only nonpositive-power terms really matter.
Recall that the lower half of the reduced Poincar\'{e} polynomial
$D_k(x)$ is determined entirely by the ambient space $X_k$ cohomology,
such that
\begin{equation}
F^-_k(x)\equiv x^{-d_k}D_k(x)\biggl\vert_{\rm nonpositive}
= \left(x^{-d_k}\prod_{i\neq k}\frac{1-x^{2a_i}}{1-x^2}\right)\Biggl\vert_{\rm nonpositive}\ .
\end{equation}
Note that positive-power terms in $x^{-d_k}D_k(x)$ are mirror
images of negative-power ones, so $F^-_k(x)$ contains the same
information as $D_k(x)$. Similarly $\Omega_{\rm Coulomb}^{(k)}(y)$
is symmetric under $y\rightarrow 1/y$, since powers of $y$ are
related to helicities, and ultimately to $SU(2)_J$ representations.

The proposed equality, $\Omega_{\rm Coulomb}^{(k)}(y)=(-y)^{-d_k}D_k(-y)$,
is then equivalent to the statement that
\begin{equation}
\Delta_k(y)\equiv \Omega_{\rm Coulomb}^{(k)}(y)-F^-_k(-y)
\end{equation}
contains only positive-power terms of $y$. In other words,
$\Delta_k(0)$ should be well-defined and equal to zero. In turn,
the latter is equivalent to the condition that
\begin{equation}\label{negative}
(1-y^2)^n\cdot\Delta_k(y)=
(1-y^2)^n\cdot\Omega_{\rm Coulomb}^{(k)}(y)
- \left((-y)^{-d_k}\prod_{i\neq k}\left(1-y^{2a_i}\right)\right)
\Biggl\vert_{\rm nonpositive}
\end{equation}
vanishes at $y=0$ and thus contains only positive powers. Thus, with
$\Omega_{\rm Coulomb}^{(k)}$ known, this would serve as a very
economical method to prove the first conjecture. Following
actual computations of $\Omega_{\rm Coulomb}$, one can see that
in general it has the form \cite{Manschot:2011xc},
\begin{equation}\label{h}
(-1)^{-n+\sum\limits_{i=1}^{n+1}a_i} \left[\frac{h(y)+(-1)^n h(y^{-1})}{(y-y^{-1})^n} \right]\ ,
\end{equation}
where the polynomial $h$ is of definite parity.
The condition that Eq.~(\ref{negative}) contains only positive-power terms demands
that non-positive power terms in $y^{n}h(y^{-1})$ should be in a very particular
form determined by the second piece on the right hand side of Eq.~(\ref{negative})
if the first conjecture is to hold. We will prove this assertion in section~4.2.

Of the polynomial $h$, only terms with power $\ge n$ enter this
comparison, so $\Delta_k(0)=0$ is clearly a necessary condition to establish
the equivalence of the two indices. Not so immediately obvious fact
is that this comparison is in fact also sufficient. How can that be?
The answer is found in the denominator in $\Omega_{\rm Coulomb} $,
which includes a factor $(y-1)^n$. Unless the numerator also has
an $n$-th order zero at $y=1$, $\Omega_{\rm Coulomb}(y)$ would be
divergent at $y=1$, which is physically unacceptable:
$\Omega_{\rm Coulomb}(1)$ is   the ground state degeneracy.
Splitting the polynomial by the power of $y$ at degree $n$,
\begin{equation}
h(y)=h_\ge(y)+h_<(y) \ ,
\end{equation}
the lower part $h_<(y)$, a polynomial of degree less than $n$, is uniquely
fixed, once $h_\ge$ is given, by the finiteness of $\Omega_{\rm Coulomb}(1)$.
On the other hand, $y^nh_\ge(y^{-1})$ is precisely the part to be
compared with the non-positive
powers of the Higgs phase expression, $(-y)^{-d_k}D_k(-y)$. Since
$(-y)^{-d_k}D_k(-y)$ is manifestly regular at $y=1$ as well, exact matching
of $y^n h_\ge(y^{-1})$ against its Higgs phase counterpart suffices to establish
$\Omega_{\rm Coulomb}^{(k)}(y) = (-y)^{-d_k}D_k(-y)$.

\subsection{Coulomb Phase Equivariant Index }

For this, let us first evaluate
$\Omega_{\rm Coulomb}^{(k)}(y)$ directly in the Coulomb phase.
Recall that for each branch of the Higgs phase,
there is a Coulomb phase counterpart. The moduli space
in Coulomb phase is spanned by solutions to multi-particle
equilibrium conditions, which, for cyclic $(n+1)$-gon quivers, are
\begin{eqnarray}\label{CoulombEQ}
\frac{a_{n+1}}{|\vec x_{n+1}-\vec x_1|}-
\frac{a_{1}}{|\vec x_{1}-\vec x_2|} &=&\zeta_1\ ,\\ \nonumber
\frac{a_{1}}{|\vec x_{1}-\vec x_2|}
-\frac{a_2}{|\vec x_{2}-\vec x_3|} &=&\zeta_2\ ,\\ \nonumber
&\vdots& \\ \nonumber
\frac{a_{n}}{|\vec x_{n}-\vec x_{n+1}|}
-\frac{a_{n+1}}{|\vec x_{n+1}-\vec x_1|} &=&\zeta_{n+1} \ .
\end{eqnarray}
Three-vectors $\vec x_i$ represent position of $i$-th charge
center. This defines a subspace of $\mathbb{R}^{3(n+1)}$.
How one obtains the Coulomb phase (equivariant) index from this,
for arbitrary number of charge centers, is a long and complicated story
\cite{Manschot:2010qz,Manschot:2011xc,Kim:2011sc}.

In particular,  Refs.~\cite{Manschot:2010qz,Manschot:2011xc} developed
an extensive technique to compute equivariant indices via
a localization method, which is in the end dictated
by the fixed point theorem associated with ${\cal J}_3$;
The fixed points are solutions
to (\ref{CoulombEQ}) with all the charge centers located
along the $z$-axis. Taking $z_1=0$ to fix the free
center of mass position, we find the Coulomb phase
equivariant index \cite{Manschot:2010qz} for our quivers
\begin{eqnarray}\label{C0}
\Omega_{\rm Coulomb}^{(k)}(y)&=&
\frac{(-1)^{\sum\limits_{i=1}^{n+1} a_i-n}}{(y-y^{-1})^n}
\left[\left(\sum_{p}s(p)\;y^{\sum\limits_{i=1}^{n+1}
a_i \;{\rm sign}[z_i-z_{i+1}]} \right)
+H_k(y)+(-1)^nH_k(y^{-1})\right]\,,
\cr \cr \cr
s(p)&=&{\rm sign}[{\rm det}\,M]\,,
\end{eqnarray}
where the summation in the round brackets is taken over all possible fixed points $p$
and $M$ is an $n \times n$ matrix with the following non-vanishing components:
\begin{eqnarray}
&&M_{i,i}= a_{i}\frac{z_{i}-z_{i+1}}{|z_i-z_{i+1}|^3}
+a_{i+1}\frac{z_{i+1}-z_{i+2}}{|z_{i+1}-z_{i+2}|^3}\,,
\cr&&M_{i,i+1}=M_{i+1,i}=-a_{i+1}\frac{z_{i+1}-z_{i+2}}{|z_{i+1}-z_{i+2}|^3}\,.
\end{eqnarray}
Note that if we have a fixed point with the ordering of charged centers
along $z$-axis as $\{\sigma(1),\sigma(2),\cdots,\sigma(n+1)\}$, then correspondingly
we will get a fixed point with the reversed ordering of charged centers
$\{\sigma(n+1),\cdots,\sigma(2),\sigma(1)\}$ by taking all $z_i$ to $-z_i$.
Therefore, we can express the contributions from the fixed points as
\begin{equation}
\sum_{p}s(p)\;y^{\sum\limits_{i=1}^{n+1} a_i \;{\rm sign}[z_i-z_{i+1}]}=G_k(y)+(-1)^n G_k(y^{-1})\,,
\end{equation}
where $G_k(y)$ is a polynomial. Thus, the polynomial $h$ in (\ref{h})
is precisely $h=G_k+H_k$ in branch $k$.

The $H_k$ terms are added by hand,\footnote{Backwardly, our first
conjecture which holds with this prescription by Manschot et. al. \cite{Manschot:2011xc}
bolsters the latter as a sensible choice.} when the quiver admits so-called
scaling solutions, when $G(y)+(-1)^nG(y^{-1})$ alone does not have
enough zero at $y=1$ and naively leads to divergent $\Omega_{\rm Coulomb}(1)$.
The prescription of Ref.~\cite{Manschot:2011xc} demands a
canceling polynomial  of the form
\begin{equation}
H_k(y)=\sum_{0\leq l<n \atop l-\sum_{i=1}^{n+1} a_i\in2{\mathbb{Z}}}\,\lambda_l\,y^l \ ,
\end{equation}
where the coefficients $\lambda_l$ are decided uniquely by
requiring that $\Omega_{\rm Coulomb}^{(k)}(y)$ is finite when $y=1$.
Note that the parity of $H_k(y)$ is required to be the same as that of
$G_k(y)$; this condition follows from the observation that $G_k$ is
of definite parity, for any given quiver, combined with the physical
requirement that $\Omega_{\rm Coulomb}^{(k)}(y)$ itself should have
the form $f_k(y)+f_k(y^{-1})$ for some polynomial $f_k$. These observations
suffice to establish the uniqueness of the canceling polynomial
$H_k$ in response to the computed $G_k$.

In order to get an explicit expression of the index for a
given quiver, we need first to solve the fixed point equations.
For quivers without closed loops, general rules for
enumerating fixed points was given in Ref.~\cite{Sen:2011aa}.
However, there is no known systematical method. For more general
quivers, let us observe that the index is invariant within each
branch, so that we may pick a particularly convenient set
of FI constants and simplify the problem. Recalling that the
$k$-th branch is described by
\begin{equation}
\sum_{i= I}^{k} \zeta_{i}> 0\ ,\quad \sum_{i= k+1}^{ J}\zeta_i<0\,,
\end{equation}
we may compute the index at the following special values of FI
constants,
\begin{equation}
\zeta_{k}=-\zeta_{k+1}>0\,,~~~\zeta_i=0~~~(i\neq k,k+1) \,,
\end{equation}
where  solutions must obey
\begin{eqnarray}
&&|z_k-z_{k+1}|=\frac{a_k}{\rho}\,,~~~|z_i-z_{i+1}|
=\frac{a_i}{\rho+\zeta_{k}}~~~(i\neq k) \,,
\cr&&\sum_{i\neq k}{\rm sign}[z_i-z_{i+1}]\frac{a_i}{\rho+\zeta_{k}}
+{\rm sign}[z_k-z_{k+1}]\frac{a_k}{\rho}=\sum_i(z_i-z_{i+1})=0 \ .
\end{eqnarray}
For each solution, ${\rm det}M$ is 
\begin{equation}
\frac{(\rho+\zeta_{k})^{2(n-1)}}
{\prod_ia_i\;{\rm sign}[z_i-z_{i+1}]}
\Big({\rm sign}[z_k-z_{k+1}]\cdot{a_k}(\rho+\zeta_{k})^2
+\sum_{i\neq k}{\rm sign}[z_i-z_{i+1}]\cdot{a_i}{\rho}^2\Big)\ .
\end{equation}
Since $\rho+\zeta_{k}>\rho>0$, if the linking numbers satisfy
the inequality $\sum_{i\neq k} a_i { t}_i<a_k$ with ${ t}_i= \pm 1$,
there is no fixed point.
On the other hand, if
$\sum_{i\neq k} a_i { t}_i\geq a_k$ holds, one  finds
a  solution
for which $s(p)$ is
\begin{equation}
s(p)=\prod_{i\neq k}t_i\,.
\end{equation}
Therefore, for cyclic $(n+1)$-gon quivers, we find
\begin{equation}\label{GC}
G_k(y)=\sum_{\{t_{i\neq k}=\pm1\}}\Big[\prod_{i\neq k}t_i\Big] \cdot
\,\Theta\Big(\sum_{i\neq k} a_i { t}_i- a_k\Big)\,
y^{\sum\limits_{i\neq k} a_i { t}_i- a_k}\,,
\end{equation}
where
\begin{eqnarray}
 \Theta(x) =
 \begin{cases}
 \hbox{1~~~~~for $x\geq 0$}\cr \hbox{0~~~~~for $x<0$}
 \end{cases}\ .
\end{eqnarray}
The Coulomb equivariant index for each branch can thus be expressed explicitly,
once the subtraction term $H_k$ is (uniquely) determined as aforementioned.

\subsection{$ \Omega_{\rm Coulomb}^{(k)}(y)=(-y)^{-d_k}D_k(-y) $ }

We are now ready to prove the first conjecture for all cyclic
Abelian quivers; by comparing the two respective routines
(\ref{C0}) and (\ref{PXM}), we will show that $\Omega_{\rm Coulomb}^{(k)}(y)$ of
the Coulomb phase equals $(-y)^{-d_k}D_k(-y)$ of the Higgs phase.

To show this, we will start by rewriting $(-y)^{-d_k}D_k(-y)$
to resemble the Coulomb side. Defining a polynomial $\tilde G_k$ as
\begin{equation}\label{tG}
y^n \tilde G_k(y^{-1})\equiv
\left(y^{-d_k}\prod_{i\neq k}\left(1-y^{2a_i}\right)\right)\Biggl\vert_{\rm nonpositive}   \ ,
\end{equation}
we have,
\begin{eqnarray}\label{DGH}
&&(-y)^{-d_k}D_k(-y)\cr\cr
&=&(-1)^{\sum\limits_{i=1}^{n+1}a_i-n}\left[\frac{\tilde G_k(y)+(-1)^n\tilde G_k(y^{-1})
+\tilde H_k(y)+(-1)^n\tilde H_k(y^{-1})}{(y-y^{-1})^n}\right]\ ,
\end{eqnarray}
where we used $(-1)^{d_k}=(-1)^{\sum_{i=1}^{n+1}a_i-n}$.
Unlike the Coulomb phase, the polynomial, $\tilde H_k$, is already fixed
by (\ref{PXM}). Nevertheless, it can be thought of as the Higgs
phase analog of $H_k$, with its highest power less than $n$, in the
sense that, if we drops it, (\ref{DGH}) would generally become
divergent at $y=1$.
Comparing this against the general expression on the Coulomb side,
\begin{eqnarray}
&&\Omega_{\rm Coulomb}^{(k)}(y)\cr\cr
&=&(-1)^{\sum\limits_{i=1}^{n+1}a_i-n}\left[\frac{G_k(y)+(-1)^nG_k(y^{-1})
+H_k(y)+(-1)^nH_k(y^{-1})}{(y-y^{-1})^n}\right]\,,
\end{eqnarray}
and, remembering earlier discussion  at the top of the section,
we see that the first conjecture holds if and only if the equality
\begin{eqnarray}\label{half}
y^n G_k(y^{-1})\Biggl\vert_{\rm nonpositive}=y^n\tilde G_k(y^{-1}) \
\end{eqnarray}
holds.

Recall from the previous subsection that uniqueness
of $H_k$, given $G_k$, was guaranteed
by three requirements: regularity of index at $y=1$, definite parity
of $G_k$, and parity of $H_k$ coinciding with that of $G_k$. All three
requirements apply to $\tilde G_k$, $\tilde H_k$ pair, in
fact trivially since $D_k(x)$ is always an even polynomial. This
means that, with only $\tilde G_k$ given as (\ref{tG}), one could
have recovered $\tilde H_k$ indirectly by writing,
\begin{equation}
\tilde H_k(y)=\sum_{0\leq l<n \atop l-\sum_{i=1}^{n+1} a_i\in2{\mathbb{Z}}}\,\tilde \lambda_l\,y^l \ ,
\end{equation}
and fixing $\tilde \lambda$'s uniquely by demanding the
regularity. Thanks to the
uniqueness, the resulting polynomial $\tilde G_k+\tilde H_k$
would be exactly the same as what we finds from the known $D_k(x)$ in
(\ref{PXM}) via (\ref{DGH}). This implies that (\ref{half})
translates to
\begin{equation}
G_k(y)+H_k(y)=\tilde G_k(y)+\tilde H_k(y) \ ,
\end{equation}
and thus proves $ \Omega_{\rm Coulomb}^{(k)}(y)=(-y)^{-d_k}D_k(-y) $.

It only remains to check Eq.~(\ref{half}). Replacing
$d_k$ by $\sum_{i\neq k} a_i -a_k-n$ in (\ref{tG}) and expanding the product,
we easily see that possible exponents are $n+a_k-\sum_{i\neq k}t_ia_i$ with
$t_i=\pm 1$; $t_i=-1$ corresponds to a multiplicative factor $-y^{2a_i}$
from the expansion, and this is accompanied by $-1$ prefactor, represented
conveniently by $t_i$ itself.  This gives
\begin{eqnarray}
y^n\tilde G_k(y^{-1})&=&\sum_{\{t_{i\neq k}=\pm1\}}\Big[\prod_{i\neq k}t_i\Big] \cdot
\,\Theta\Big( \sum_{i\neq k} a_i { t}_i-a_k-n\Big)\,
y^{-\sum\limits_{i\neq k} a_i { t}_i+ a_k+n}\ ,
\end{eqnarray}
whose features are reminiscent of the Coulomb phase expression in Eq.~(\ref{GC}).
Indeed, starting with the latter and flipping $y\rightarrow 1/y$,  we find
\begin{eqnarray}
y^n G_k(y^{-1})\Biggl\vert_{\rm nonpositive}&=&y^n\sum_{\{t_{i\neq k}=\pm1\}}\Big[\prod_{i\neq k}t_i\Big] \cdot
\,\Theta\Big(\sum_{i\neq k} a_i { t}_i- a_k\Big)\,
y^{-\sum\limits_{i\neq k} a_i { t}_i+ a_k}\Biggl\vert_{\rm nonpositive} \\\cr\cr
&=&\sum_{\{t_{i\neq k}=\pm1\}}\Big[\prod_{i\neq k}t_i\Big] \cdot
\,\Theta\Big( \sum_{i\neq k} a_i { t}_i-a_k-n\Big)\,
y^{-\sum\limits_{i\neq k} a_i { t}_i+ a_k+n}\ .\nonumber
\end{eqnarray}
This gives $(-y)^{-d_k}D_k(-y) = \Omega_{\rm Coulomb}^{(k)}(y)$ as promised,
and concludes the proof of the first conjecture for all cyclic Abelian quivers.

\subsection{Examples}

The procedure we adopted to prove the first conjecture
for cyclic Abelian quivers offers an interesting and
perhaps more economical method to find $\Omega_{\rm Coulomb}(y)$
without ever going into the Coulomb phase, by computing
$\tilde G(y)+\tilde H(y)$ of the Higgs phase instead of
$G(y)+H(y)$ of the Coulomb phase. With $\tilde G_k(y)$ given
by
\begin{equation}\label{again}
y^n \tilde G_k(y^{-1})=
\left(y^{-d_k}\prod_{i\neq k}\left(1-y^{2a_i}\right)\right)\Biggl\vert_{\rm nonpositive}   \ ,
\end{equation}
the simplicity comes from two aspects. First, the right hand side is
entirely determined by the ambient manifold $X_k$;  the complicated
F-term conditions enter only via a single integer, $d_k$.
Second,  $\prod_{i\neq k}\left(1-y^{2a_i}\right)=(1-y^2)^nP[X_k](-y)$
contains exactly the same information as  $P[X_k](x)$,
yet is far less cluttered. Thanks to these features, and the
respective uniqueness of $H_k$ and $\tilde H_k$, we can compute
 the Coulomb phase equivariant index
\begin{equation}
\Omega_{\rm Coulomb}^{(k)}(y)\quad\leftarrow\quad
(-1)^{\sum  a_i -n}\times \frac{h_k(y)+(-1)^nh_k(y^{-1})}{(y-y^{-1})^n}
\end{equation}
itself, with $h_k(y)=\tilde G_k(y)+\tilde H_k(y)$.
Below, we exploit this for general 3-gon and 4-gon examples, and
display the equivariant indices in their full generality.  We suspect that
similar simplification will occur for more general class of quivers.

\subsubsection*{3-Gons}

Without loss of generality, let us consider the third branch and
compute $\tilde G_3$ using the Higgs phase picture. There are
three cases in all.

\begin{itemize}
\item
Case 1: $a_3>a_1+a_2-2$

Because  the relevant Higgs phase does not exist, we have $\tilde G_3(y^{-1})=0$, which also
means that $\Omega_{\rm Coulomb}^{(3)}(y)=0$.

\item
Case 2: $a_{\sigma(1)}\geq a_{\sigma(2)}+a_3+2$
where $\sigma$ denotes a permutation of $\{1,2\}$.

In this case, two terms show up in Eq.~\eqref{tG} for $\tilde G_3$
\begin{eqnarray}
\tilde G_3(y^{-1})&=&y^{-a_{\sigma(1)}-a_{\sigma(2)}+a_3}
-y^{-a_{\sigma(1)}+a_{\sigma(2)}+a_3}
\,.
\end{eqnarray}
We also notice that no counter terms are needed, $\tilde H_3(y)=0$.

\item
Case 3: The twisted $3$-gon condition is obeyed
\begin{eqnarray}
a_{1}&<& a_{2}+a_{3}+2\ ,
\cr
a_{2}&<& a_{1}+a_{3}+2 \ ,
\cr
a_{3}&\leq& a_{1}+a_{2}-2 \ ,
\end{eqnarray}
whereby only one term exists in $\tilde G_3$,
\begin{eqnarray}
\tilde G_3(y^{-1})=y^{-a_1-a_2+a_3} \ ,
\end{eqnarray}
and the counter terms are easily determined as
\begin{eqnarray}
&\tilde H_3(y)=
\begin{cases}
-y\,,& {\rm if~}{\sum_ia_i\rm{~is~odd \ ;}}\\
-1\,,& {\rm if~}{\sum_ia_i~\rm is~even}\ .
\end{cases}
\end{eqnarray}

\end{itemize}

\subsubsection*{4-Gons}

Again without loss of generality, let us consider the
fourth branch and determine $\tilde G_4$.
There are six cases in all.

\begin{itemize}

\item
Case 1: $a_4>a_1+a_2+a_3-3$

The corresponding Higgs phase is null, so we find $\tilde G_4(y^{-1})=0$ and
thus $\Omega_{\rm Coulomb}^{(4)}(y)=0$.

\item
Case 2: $a_{\sigma(1)}\geq a_{\sigma(2)}+a_{\sigma(3)}+a_4+3$
where $\sigma$ is a permutation of $\{1,2,3\}$.

Following the same procedure, we get from Eq.~(\ref{tG})
\begin{eqnarray}
\tilde G_4(y^{-1})&=&y^{-a_{\sigma(1)}-a_{\sigma(2)}-a_{\sigma(3)}+a_4}
-y^{-a_{\sigma(1)}+a_{\sigma(2)}-a_{\sigma(3)}+a_4}
\cr&&-y^{-a_{\sigma(1)}-a_{\sigma(2)}+a_{\sigma(3)}+a_4}
+y^{-a_{\sigma(1)}+a_{\sigma(2)}+a_{\sigma(3)}+a_4}\, ,
\end{eqnarray}
which requires no counter terms; $\tilde H_4(y)=0$.

\item
Case 3: The twisted $4$-gon condition is obeyed
\begin{eqnarray}
a_{1}&<& a_{2}+a_{3}+a_{4}+3 \ ,
\cr
a_{2}&<& a_{1}+a_{3}+a_{4}+3\ ,
\cr
a_{3}&<& a_{1}+a_{2}+a_{4}+3\ ,
\cr
a_{4}&\leq& a_{1}+a_{2}+a_{3}-3\ ,
\end{eqnarray}
and in addition
\begin{eqnarray}
a_1+a_2&\geq&a_3+a_4+3\ ,\cr
a_1+a_3&\geq&a_2+a_4+3\ ,\cr
a_2+a_3&\geq&a_1+a_4+3\ .
\end{eqnarray}
We find
\begin{eqnarray}
\tilde G_4(y^{-1})&=&y^{-a_1-a_2-a_3+a_4}
-y^{a_1-a_2-a_3+a_4}
\cr&&-y^{-a_1+a_2-a_3+a_4}-y^{-a_1-a_2+a_3+a_4}\ ,
\end{eqnarray}
with the counter terms
\begin{eqnarray}
&\tilde H_4(y)=
\begin{cases}
-2a_4y\,,& {\rm if~}\sum_ia_i\rm{~is~odd\ ;}\\
-a_4y^2\,,&{\rm if~}\sum_ia_i{~\rm is~even}\ .
\end{cases}
\end{eqnarray}

\item
Case 4: The twisted $4$-gon condition
is obeyed and in addition
\begin{eqnarray}
a_{\sigma(1)}+a_{\sigma(2)}&\geq&a_{\sigma(3)}+a_4+3\ ,\cr
a_{\sigma(1)}+a_{\sigma(3)}&\geq&a_{\sigma(2)}+a_4+3\ ,\cr
a_{\sigma(2)}+a_{\sigma(3)}&<&a_{\sigma(1)}+a_4+3\ .
\end{eqnarray}
Similarly,
\begin{eqnarray}
\tilde G_4(y^{-1})&=&y^{-a_{\sigma(1)}-a_{\sigma(2)}-a_{\sigma(3)}+a_4}
-y^{-a_{\sigma(1)}+a_{\sigma(2)}-a_{\sigma(3)}+a_4}
\cr&&-y^{-a_{\sigma(1)}
-a_{\sigma(2)}+a_{\sigma(3)}+a_4}\ ,
\end{eqnarray}
with the counter terms
\begin{eqnarray}
&\tilde H_4(y)=
\begin{cases}
(a_{\sigma(1)}-a_{\sigma(2)}-a_{\sigma(3)}-a_4)y\,,&   {\rm if~}\sum_ia_i\rm{~is~odd\ ;} \\
\frac{1}{2}(a_{\sigma(1)}-a_{\sigma(2)}-a_{\sigma(3)}-a_4)y^2\,,&{\rm if~}\sum_ia_i{~\rm is~even}\ .
\end{cases}
\end{eqnarray}

\item
Case 5: The twisted $4$-gon condition
is satisfied and in addition
\begin{eqnarray}
a_{\sigma(1)}+a_{\sigma(2)}&\geq&a_{\sigma(3)}+a_4+3\ ,\cr
a_{\sigma(1)}+a_{\sigma(3)}&<&a_{\sigma(2)}+a_4+3\ ,\cr
a_{\sigma(2)}+a_{\sigma(3)}&<&a_{\sigma(1)}+a_4+3\ .
\end{eqnarray}
In this case,
\begin{eqnarray}
\tilde G_4(y^{-1})=y^{-a_{\sigma(1)}-a_{\sigma(2)}-a_{\sigma(3)}+a_4}
-y^{-a_{\sigma(1)}-a_{\sigma(2)}+a_{\sigma(3)}+a_4}\ ,
\end{eqnarray}
with the counter terms
\begin{eqnarray}
&\tilde H_4(y)=
\begin{cases}
-2a_{\sigma(3)}y\,,&  {\rm if~}\sum_ia_i\rm{~is~odd\ ;} \\
-a_{\sigma(3)}y^2\,,&{\rm if~}\sum_ia_i{~\rm is~even}\ .
\end{cases}
\end{eqnarray}

\item
Case 6: The twisted $4$-gon condition
is satisfied and in addition
\begin{eqnarray}
a_1+a_2&<&a_3+a_4+3\ ,\cr
a_1+a_3&<&a_2+a_4+3\ ,\cr
a_2+a_3&<&a_1+a_4+3\ .
\end{eqnarray}
In this last case,
\begin{eqnarray}
\tilde G_4(y^{-1})=y^{-a_1-a_2-a_3+a_4}\ ,
\end{eqnarray}
and the counter terms are
\begin{eqnarray}
&\tilde H_4(y)=\begin{cases}
(-a_1-a_2-a_3+a_4)y\,,& {\rm if~}\sum_ia_i\rm{~is~odd \ ;} \\
\frac{1}{2}(-a_1-a_2-a_3+a_4)y^2\,,&{\rm if~}\sum_ia_i{~\rm is~even}\ .
\end{cases}
\end{eqnarray}

\end{itemize}

\section{Enumerating Cohomology}

Having proved the first conjecture for all cyclic Abelian quivers,
we now turn to the Higgs phase cohomology. Two aspects are of main concern
in this section. The first is the proposed invariance of
$H(M_k)/i^*_{M_k}(H(X_k))$. While more detailed information
on the matter will be offered in the next section, including
the general proof of the second conjecture at fully refined level
and a routine for determining the full Hodge-decomposed
cohomologies, we offer here a more preliminary, perhaps more
intuitive view on the conjecture. The relatively simple ambient
space $X_k$ allows a pictorial realization of $i^*_{M_k}(H(X_k)) $
via an $n$-dimensional triangular lattice, which produces
a simple and intuitive proof of the invariance, albeit
only at the level of the numerical index. The second is an explicit
and elementary counting formula for $D_k(-1)={\rm dim}[i^*_{M_k}(H(X_k))]$,
or equivalently the counting of the Coulomb phase ground states,
which also follows from the same lattice realization of $i^*_{M_k}(H(X_k))$.

The usual numerical index for the Higgs phase is the Euler number  $\chi(M_k)$,
which can be computed as an integral of the top Chern class.
Note that the total Chern class of $M_k$ is given by
\begin{equation}
c(M_{k})=\frac{\prod_{i\neq k} (1+J_i)^{a_i}}{\left(1+\sum_{i\neq k} J_i\right)^{a_{k}}}\,,
\end{equation}
where $J_i$ is the K\"ahler form of the ${\mathbb{CP}}^{a_i-1}$ factor
and the sum and the product over $i$ run from $1$ to $n+1$, except $i=k$.
Then the evaluation of $\chi(M_{k})$ follows
\begin{equation}\label{Euler}
\chi(M_{k})=
\int_{X_k}\frac{ \prod_{i\neq k} (1+J_i)^{a_i} }{\left(1+\sum_{i\neq k} J_i\right)^{a_{k}}} \cdot
\Big(\sum_{i\neq k} J_i \Big)^{a_k}\,,
\end{equation}
where we are supposed to extract the coefficient of
$ \prod_{i\neq k} J_i^{a_i-1}$ of the integrand.
Alternatively
and more explicitly, this Euler number can also be expressed as
\begin{eqnarray}\label{Euler2}
\chi(M_{k})&=&\prod_{i\neq k}a_i-\int_0^\infty ds\;
\Big(e^{-s}\prod_{i=1}^{n+1} L_{a_i-1}^1(s) \Big)
\end{eqnarray}
with Laguerre polynomials $L_{a_i-1}^1(s)$.\footnote{See Appendix E of Ref.~\cite{Denef:2007vg} for
$n=2$ example. Generalization to $n>2$  is straightforward.}
This last expression~\eqref{Euler2} greatly simplifies the proof since the variation of
the Euler number across walls, i.e. between adjacent branches,
is only due to the first term. Thus $k$-independence of $\chi(M_k)-D_k(-1)$ is
equivalent to the $k$-independence of
\begin{equation}\label{kindep}
E_k\equiv \prod_{i\neq k}a_i- D_k(-1) \ ,
\end{equation}
and hence, it is enough to show that $E_k=E_{k'}$ for any $k,k'$ of a given cyclic quiver.
We shall shortly reduce this to a counting problem in an $n$-dimensional
triangular lattice.

Now, evaluation of $D_k(-1)$ can be conveniently cast into
a counting of lattice points, thanks to
Eq.~(\ref{poincare-poly}).\footnote{We are indebted to HwanChul Yoo
for suggesting the lattice approach to the present counting problem,
and also for suggesting the triangular lattice, as a slanted form of
the rectangular lattice. The latter makes the invariance proof in
subsequent subsections more intuitive and manifest.}
With the Poincar\'e polynomial
\begin{equation}
\sum b_{l}(X_k) x^l=\frac{\prod_{i\neq k}(1-x^{2a_i})}{(1-x^2)^n}
=\prod_{i\neq k} \left(1+x^2+x^4+\cdots+x^{2(a_i-2)}\right)\ ,
\end{equation}
we see that the Betti number $b_{2m}(X_k)$ of the ambient space
equals the number of lattice points at level
$m$, call it ${\cal N}_k(m)$, in an $n$-dimensional rectangular hyper-cube,
bounded between 0 and $a_i-1$ for each direction $i\neq k$.
Later we will slant the lattice to make the bounded region into a
hyper-parallelogram of lengths $a_i-1$, instead of a hyper-cube.
Here, however, let us simply imagine a rectangular lattice for
the purpose of setting a combinatorial picture.
The level of a lattice point $(\nu_1,\nu_2,....,\nu_n)$ is defined as $\sum \nu_j$, so
the level spans from 0 to $\mu_k=\sum_{i\neq k} (a_i-1)=\sum_{i\neq k}{a_i}-n$.
Thus, we find that
\begin{equation}
D_k(-1)={\cal N}_k(d_k/2)+2\sum_{m< d_k/2} {\cal N}_k(m)=
\sum_{m\le  d_k/2} {\cal N}_k(m)~~+\sum_{m> \mu_k -d_k/2} {\cal N}_k(m)\ .
\end{equation}
Since the total number of lattice points in such a hyper-cube
is $\prod_{i\neq k} a_i$ by construction,
we learn immediately that $E_k$ is counted as
the number of lattice points whose level $m$ is between $d_k< m\le \mu_k-d_k/2 $
\beq\label{5.7}
E_k=\prod_{i\neq k}a_i -D_k(-1) =\sum\limits_{d_k/2 < m \le \mu_k-d_k/2} {\cal N}_k(m) \ .
\eeq
We shall prove the second conjecture by confirming that this counting is independent of $k$ altogether.

A crucial ingredient for the proof comes from the fact
that the upper bound for levels
\begin{equation}
\mu_k -d_k/2= \sum_{i\neq k}a_i -n-\frac{\sum_{i\neq k} a_i -n -a_k}{2}
=\frac{\sum_{i=1}^{n+1} a_i -n}{2}
\end{equation}
is independent of $k$.
As we shall see shortly, this allows a further transformation of the problem, once the lattice is slanted,
to the counting of lattice points in an overlapping pair of mutually-inverted
$n$-dimensional hyper-tetrahedrons (that is, $n$-simplices), of size
$\sim ({\sum_{i=1}^{n+1} (a_i -1)})/{2}$.

\subsection{3-Gons}
\begin{figure}[!h]
\centering
\includegraphics[width=8cm]{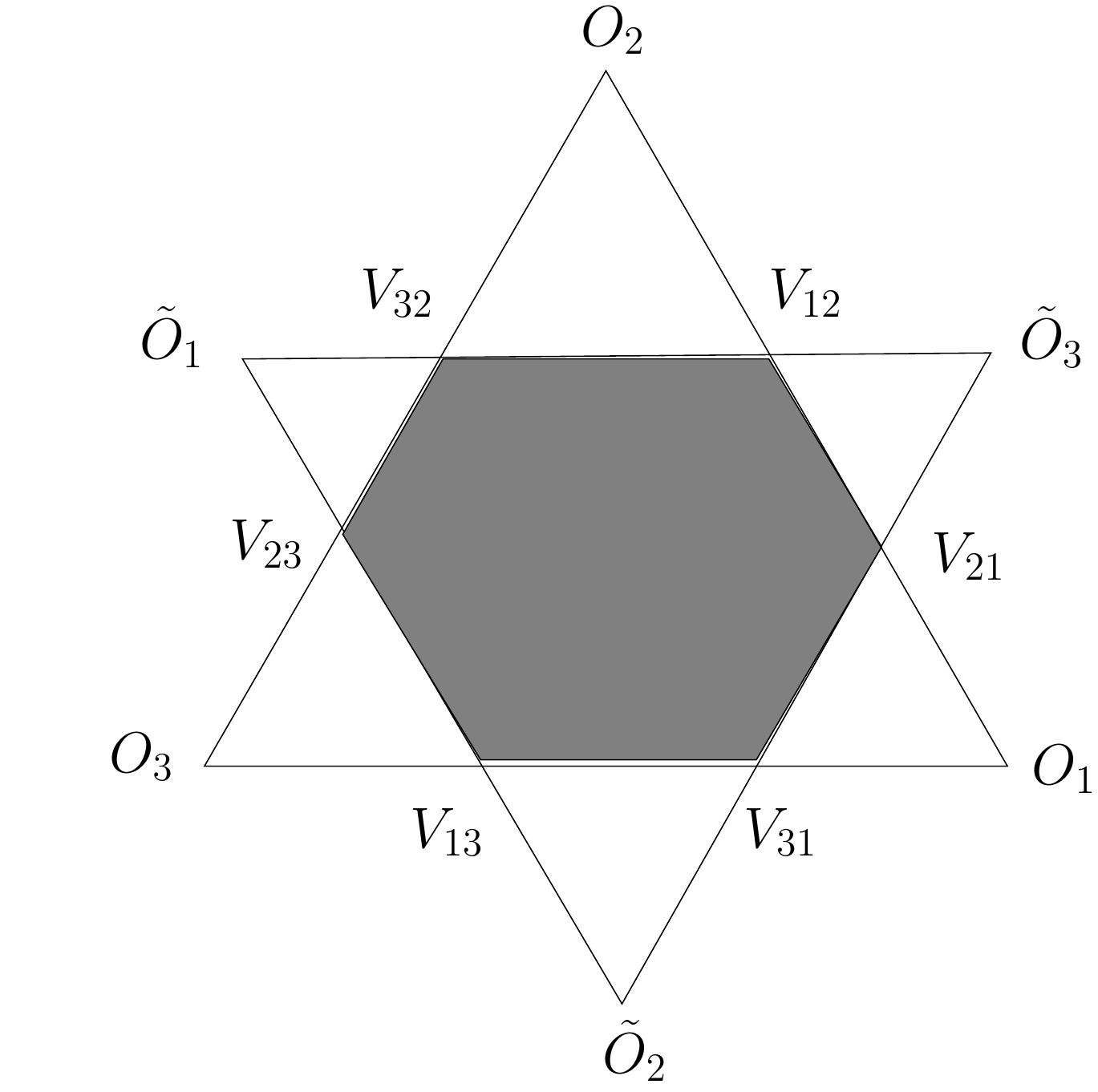}
\parbox{6in}{\caption{\small A pictorial representation of the Poincar\'{e}
polynomials for a 3-gon quiver when $d_i \geq -1$ for $i=1,2,3$.
Two mutually-inverted equilateral triangles $O_1 O_2 O_3$ and
$\tilde O_1 \tilde O_2 \tilde O_3$ are placed in a triangular
lattice, overlapping with each other at a shaded hexagonal region.
In each branch $k$, $E_k = \frac{ a_1 a_2 a_3}{a_k} - D_k(-1)$ counts the number
of lattice points inside this hexagon.  }\label{fig:3gon}}
\end{figure}

For a clear and simple picture, let us first focus on $n=2$ cases.
Once these are dealt with, generalization to higher $n$ should be straightforward.
Suppose that the linking numbers of a given quiver obey the inequalities
\begin{eqnarray}\label{firstclass}
a_1 +1 &\le & a_2+a_3\ , \nonumber \\
a_2 +1 &\le & a_3+a_1\ , \\
a_3 +1 &\le & a_1+a_2\ , \nonumber
\end{eqnarray}
or equivalently,
\beq
d_i \geq -1 \ , ~i= 1,2,3 \,,
\eeq
and let us consider the ``1-2 parallelogram'' $O_3 V_{31} \tilde{O_3} V_{32}$
in a triangular lattice (see Figure~\ref{fig:3gon}), whose two sides have
 $a_1$ and $a_2$ lattice points lying on them, respectively, {\it i.e.},
  $\overline{O_3 V_{31}} = a_1-1$ and $\overline{O_3 V_{32}}= a_2-1$.
It is easy to see that this 1-2 parallelogram naturally represents the
Poincar\'{e} polynomial of the ambient space $X_3 = \IC \IP^{a_1 -1}
\times \IC \IP^{a_2-1}$ in branch 3.
In particular, the fact that it contains total of $a_1 \cdot a_2$
lattice points inside implies that\footnote{When counting lattice
points inside a bounded region, we always include those lying on the boundary.}
\beq
a_1 \cdot a_2 ~= \sum\limits_{l=0}^{a_1 + a_2 -2} b_{2l} (X_3) ~=
\sum\limits_{l=0}^{2 (a_1 + a_2 - 2)} (-1)^l\, b_l (X_3) ~=~   \chi(X_3)\ ,
\eeq
where, in the second step, vanishing of odd Betti numbers has been used.

Now, we place two parallel hyperplanes at lattice distances
$\ce*{\frac{d_3+1}{2}}$ and $\fl*{\frac{d_3+1}{2} }$
(towards the inside region of the parallelogram), respectively,
from the vertices $O_3$ and $\tilde{O_3}$, where $\lceil \cdot \rceil$
and $\lfloor \cdot \rfloor$ are the usual ceiling and floor functions:
\bea
\nn \lceil x \rceil &=& {\rm min}\, \{ m \in \IZ \; | \;m \geq x \} \,, \\
 \nn \lfloor x \rfloor &=&{\rm max} \{ m \in \IZ\; | \;m \leq x \}  \,.
\eea
We also denote the eight lattice points by $\tilde O_2, V_{13}, V_{23},
\tilde O_1, O_1, V_{21}, V_{12}$ and $O_2$, which arise from the
intersection of these two hyperplanes with the four sides of
the 1-2 parallelogram (the reason for this naming will become
clearer when we consider general $n$ cases).
It is then straightforward to see that the quantity
$E_3 = a_1 a_2  - D_3(-1)$ counts the lattice points inside the hexagonal
region $V_{13} V_{31} V_{21} V_{12} V_{32} V_{23}$ (shaded in Figure~\ref{fig:3gon})
amongst the $a_1 a_2$ that are contained in the 1-2 parallelogram.

\begin{figure}[!h]
\centering
\includegraphics[width=8.5cm]{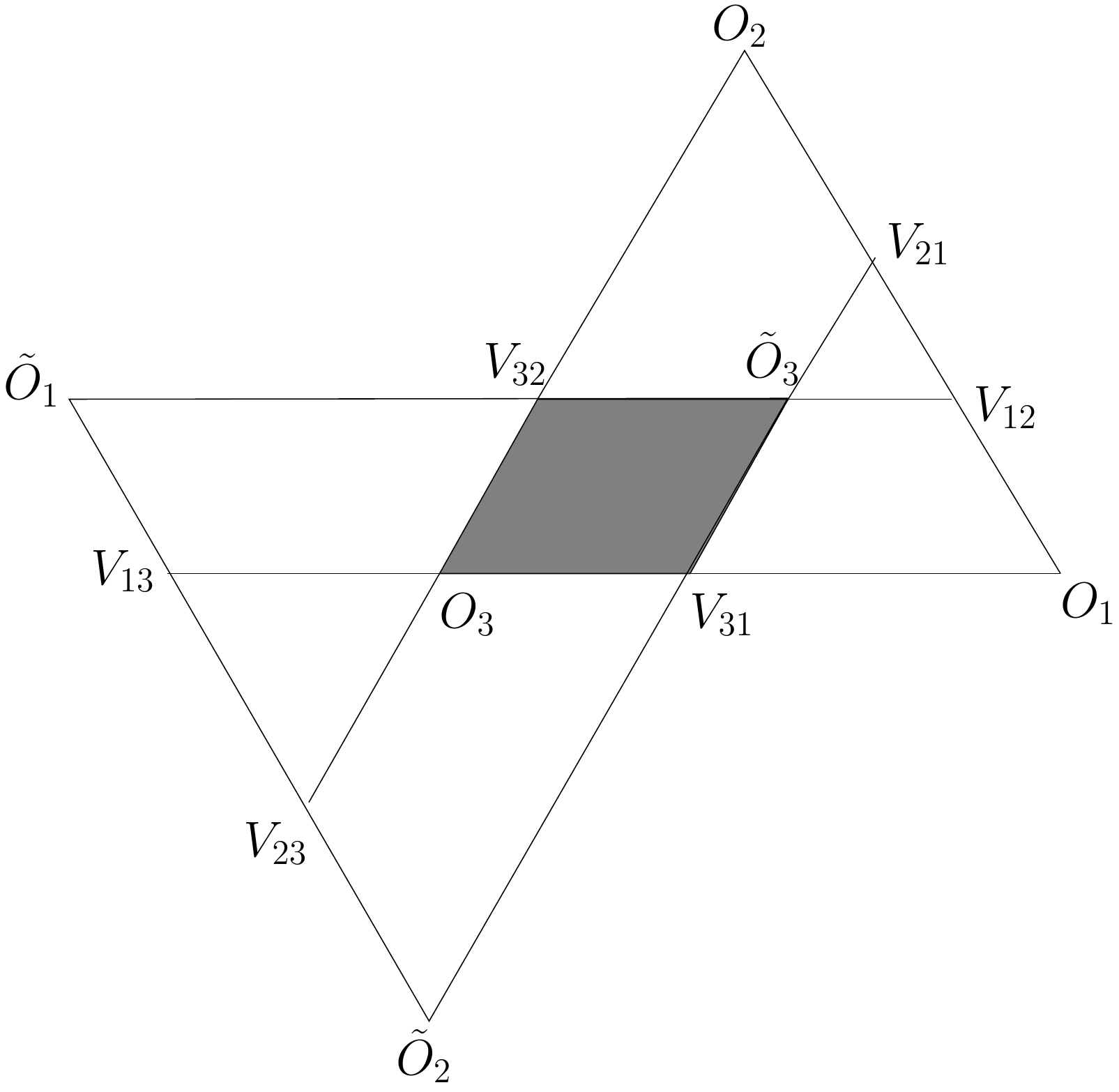}
\parbox{6in}{\caption{\small A pictorial representation of the
Poincar\'{e} polynomials for a 3-gon quiver when $d_3 < -1$.
Two mutually-inverted equilateral triangles $O_1 O_2 O_3$ and
$\tilde O_1 \tilde O_2 \tilde O_3$ are placed in a triangular lattice,
overlapping with each other at the shaded parallelogram.
In each branch $k$, $E_k = \frac{ a_1 a_2 a_3}{a_k} - D_k(-1)$ counts the number
of lattice points inside this parallelogram.
In particular, $D_3(-1)=0$ and $E_3=a_1 a_2$ in branch 3.}\label{fig:viol3gon}}
\end{figure}

We shall now show that the quantity $E_1 = a_2 a_3 - D_1(-1)$ in branch 1
 also counts the lattice points inside the same hexagon.
This can most easily be understood by exchanging the roles of $O_3$ and $O_1$.
Note first that the sides of the equilateral triangle $O_1 O_2 O_3$
have the following length:
\bea
\nn \overline{O_3 O_1} &=& \overline {O_3 V_{31} } +
\overline{V_{31} O_1}= \overline {O_3 V_{31} } + \overline{V_{31} V_{21}} \\
\nn &=& (a_1-1) + \left( (a_2-1) - \fl*{\frac{d_3+1}{2}} \right) \\
\nn &=& a_1 + a_2 -2 - \fl*{\frac{a_1 + a_ 2 - a_3 -1}{2}} \\
\nn &=& \ce*{ a_1+a_2-2 - \frac{a_1 + a_2 -a_3 -1 }{2}} \\
 &=& \ce*{\frac{a_1 + a_2 +a_3 - 3}{2}} \equiv l \,,
\eea
which is of a symmetric form under permutation of $a_i$'s.
Next, let us also note that
\beq
\overline{O_1 V_{12}} = \overline{O_3 V_{32}} = a_2-1 \,,
\eeq
and that
\bea
\nn\overline{O_1 V_{13}} &=& l - \overline{O_3 V_{13}} = l - \ce*{\frac{ d_3 +1}{2}} \\
\nn &=& \ce*{\frac{a_1 + a_2+a_3 -3}{2}} - \ce*{\frac{a_1 + a_2 - a_3 -1}{2}} \\
\nn &=& {\frac{a_1 + a_2+a_3 -3}{2}} - {\frac{a_1 + a_2 - a_3 -1}{2}} \\
 &=& a_3-1 \ , 
\eea
where the second last equality comes from the fact that $a_1 + a_2 + a_3 - 3$
and $a_1+a_2 -a_3 -1$ always have the same parity.
So we conclude that the parallelogram $O_1 V_{12} \tilde O_1 V_{13}$ forms
a ``2-3 parallelogram'' in that its two sides have $a_2$ and $a_3$
lattice points lying on them, respectively, {\it i.e.},
$\overline{O_1 V_{12}} = a_2-1$ and $\overline{O_1 V_{13}}= a_3-1$.
What still remains to be verified for complete symmetry is
that the hyperplane ${\tilde O_2 \tilde O_3}$ is placed
at lattice-distance $\ce*{\frac{d_3+1}{2}}$ from $O_1$ and the
hyperplane ${O_2 O_3}$, at $\fl*{\frac{d_3+1}{2}}$ from $\tilde O_1$.
A few lines of algebra confirm these easily:
\bea
 \overline{O_1 V_{21}} &=& (a_2-1) - \fl*{\frac{d_3+1}{2}} =
\ce*{\frac{d_1+1}{2}} \,,\\
 \overline{\tilde O_1 V_{23}} &=& \overline{ V_{23} V_{32}}
= (a_2-1) - \ce*{\frac{d_3+1}{2}} = \fl*{\frac{d_1+1}{2}}\,. 
\eea
Therefore, $E_1$ also counts the lattice points in the same
region $V_{13} V_{31} V_{21} V_{12} V_{32} V_{23}$, and hence, $E_3= E_1$.
A similar argument works in branch 2 and this completes the invariance proof
\beq
E_1= E_2 = E_3 \ ,
\eeq
in $n=2$ cases.

\paragraph{Remarks:}
\begin{itemize}
\item The hyperplane $\tilde O_1 \tilde O_2$ at distance
$\ce*{\frac{d_3+1}{2}}$ from $O_3$ is located closer to $O_3$ than
the other hyperplane $O_1 O_2$ is. This is guaranteed by
$$\ce*{\frac{d_3 + 1}{2}} = \ce*{\frac{a_1 + a_2 - a_3 - 1}{2}}
\leq \ce*{\frac{a_1 + a_2+a_3 -3}{2}} = l \,.$$
Thus, the lattice points in the hexagonal region indeed have
intermediate levels as described in Eq.~\eqref{5.7}.
\item We have assumed from the start that the linking numbers
$a_i$ obeyed the inequalities~\eqref{firstclass}. However, our
argument works in general if we consider all the
edge lengths as signed ones. Note that only one of the
three inequalities~\eqref{firstclass} can be violated at a time, and
suppose $d_3 <-1$. Then having the length $\ce*{\frac{d_3 + 1 }{2}}$
negative, for example, means that the hyperplane ${\tilde O_1 \tilde O_2}$
is located outside the 1-2 parallelogram $O_3 V_{31} \tilde O_3 V_{32}$ (see Figure~\ref{fig:viol3gon}).
It is easy to see in this case that $\tilde O_3$ sits inside
the triangle $O_1 O_2 O_3$. So the shaded region, which is
the overlap of the two triangles $O_1 O_2 O_3$ and
$\tilde O_1 \tilde O_2 \tilde O_3$, is the entire 1-2
parallelogram and hence, $E_3 = a_1 a_2$, or equivalently, $D_3(-1) = 0$.
\end{itemize}

\subsection{$(n+1)$-Gons}
\begin{figure}[!h]
\centering
\includegraphics[width=8cm]{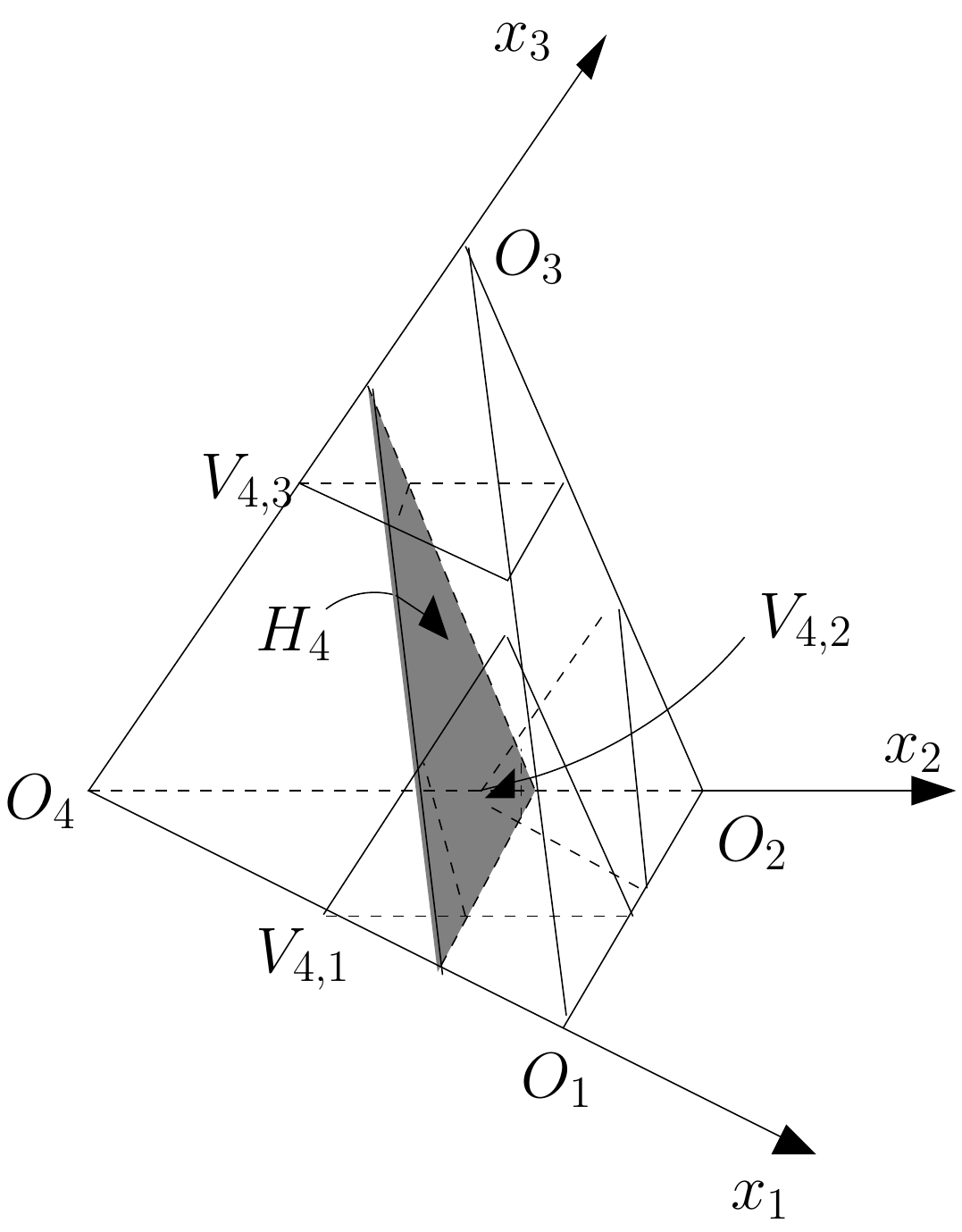}
\parbox{6in}{\caption{\small A pictorial representation
of the Poincar\'{e} polynomials for a 4-gon quiver when
 $d_i \geq -1$ for $i=1,2,3,4$. }\label{fig:tetrahedron}}
\end{figure}

Having had enough experiences with $n=2$ cases, we can
rather formally proceed to general $n$ cases.
Let $S_n = O_1 O_2 \cdots O_n O_{n+1}$ be the $n$-simplex
of edge length $$l_n\equiv \ce*{\frac{\sum\limits_{i=1}^{n+1} a_i - (n+1)}{2}}\,, $$
where $a_i$ are the $n+1$ linking numbers of a given $(n+1)$-gon quiver.
Let us now place $S_n$ in an $n$-dimensional triangular lattice
so that $O_{n+1}$ sits at the origin and the edges
$\overline{O_{n+1} O_i}$ lie along the positive $x_i$-axes for
 $i=1, 2, \cdots, n$ (see Figure~\ref{fig:tetrahedron} for an $n=3$
 case, in which we have a 3-simplex, namely, a tetrahedron).
For simplicity, let us suppose that the linking numbers obey the inequalities
\begin{eqnarray}\label{firstclass_general}
a_1 +(n-1) &\le & a_2+a_3 + \cdots + a_{n+1}\ , \nonumber \\
a_2 +(n-1) &\le & a_1+a_3+\cdots+a_{n+1}\ , \\
&\vdots& \nn \\
a_{n+1} +(n-1) &\le & a_1+a_2 + \cdots + a_n \ , \nonumber
\end{eqnarray}
or equivalently,
\beq
d_i \geq -1 \ ,~{\text{for }}i= 1,2,\cdots , n+1 \,.
\eeq

Now, we place $n+1$ hyperplanes $\CH_j$ at lattice distances
$\ce*{\frac{d_j+1}{2}}$ (towards the inside region of $S_n$)
from the vertices $O_j$.
These hyperplanes $\CH_j$ form another simplex $\tilde S_n$,
which has an inverted shape compared with the original simplex $S_n$.
Let $I \equiv S_n \cap \tilde S_n$ be the overlap of $S_n$ with $\tilde S_n$.
We claim that the number of lattice points in $I$ equals $E_k$
 in any branches, which implies that $E_k$ is independent of the branch choice, $k$.

\begin{figure}[!h]
\centering
\includegraphics[width=8cm]{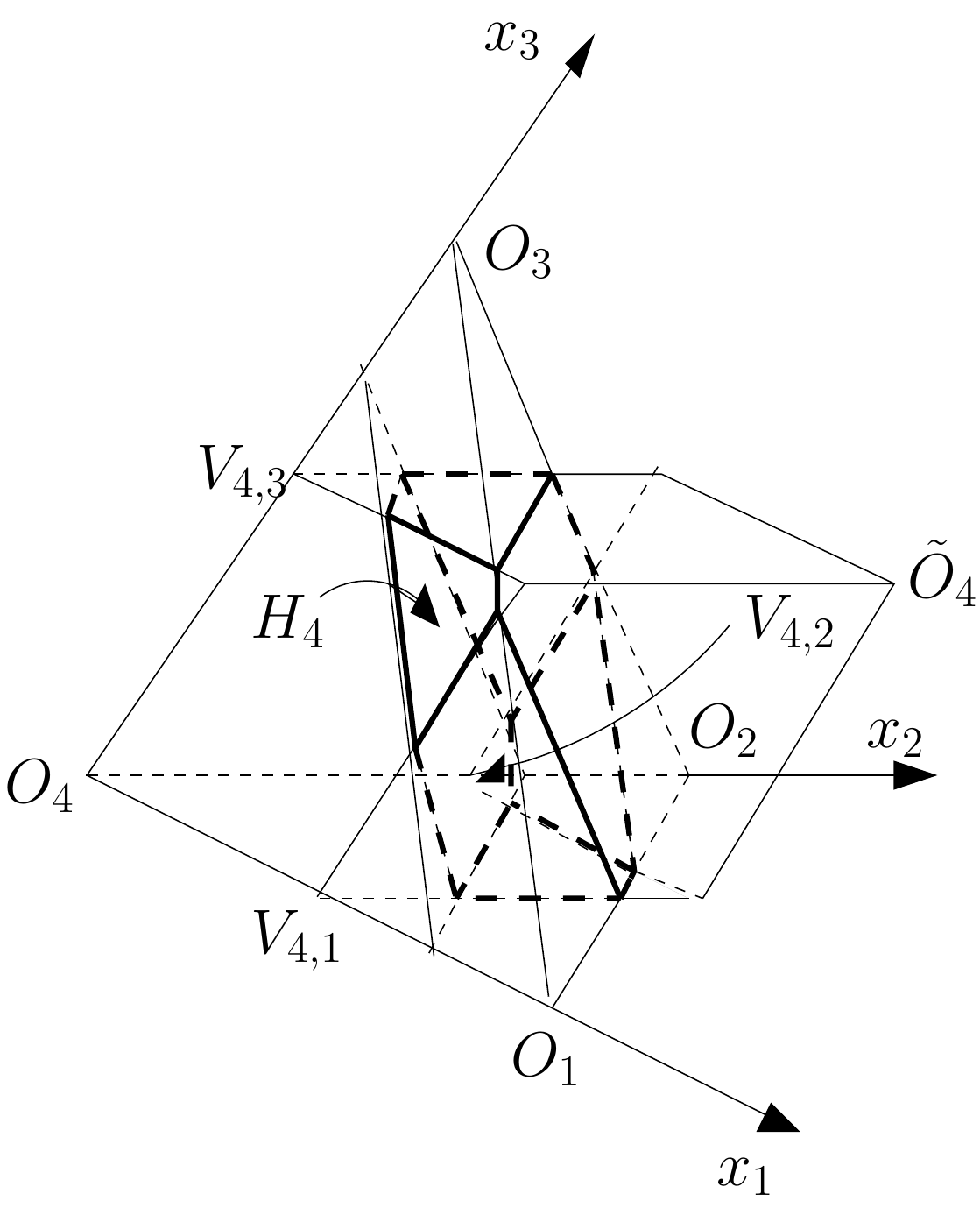}
\parbox{6in}{\caption{\small A pictorial representation of the
Poincar\'{e} polynomials for a 4-gon quiver when $d_i \geq -1$
for $i=1,2,3,4$. A hyper-parallelogram with sides of lengths
$a_i -1$  has been added to Figure~\ref{fig:tetrahedron}. Counting
of the lattice points inside the region with thick edges gives
$E_k = \frac{ a_1 a_2 a_3}{a_k} - D_k(-1)$ in any branches $k$.}\label{fig:tetrahedronE}}
\end{figure}

Thanks to the symmetry, without loss of generality, we may only consider branch $n+1$.
Let us denote by $V_{n+1,j\;}$ the $x_j$-cuts of the hyperplanes $\CH_j$ for $j=1, \cdots, n$.
It is easy to see that these $n$ hyperplanes $\CH_1, \cdots, \CH_{n}$
together with  the $n$ coordinate planes, $x_j=0$ for $j=1,\dots,n$,
 form a bounded hyper-parallelogram (see Figure~\ref{fig:tetrahedronE} for an $n=3$ example).
The $n$ sides of this object have the following lengths
\bea
\nn \overline{O_{n+1} V_{n+1, j}} &=& l_n - \ce*{\frac{d_j +1}{2}} \\
\nn &=&\ce*{\frac{\sum\limits_{i=1}^{n+1} a_i - (n+1)}{2}} -
\ce*{\frac{\sum\limits_{i=1}^{n+1} a_i - 2 a_j - n+1}{2}} \\
&=& a_j -1 \,, \label{length}
\eea
where in the last step we have used the fact that
$\sum_{i=1}^{n+1} a_i - (n+1)$ and $\sum_{i=1}^{n+1} a_i - 2 a_j - n+1$
have the same parity.
Therefore, this hyper-parallelogram naturally represents
the Poincar\'{e} polynomial of the ambient space
$X_{n+1} = \prod_{i=1}^n \IC \IP^{a_i -1}$ in branch $n+1$.

Let $\tilde O_{n+1}$ be the opposite vertex of $O_{n+1}$ in this
hyper-parallelogram.
According to Eq.~\eqref{length}, this vertex $\tilde O_{n+1}$
has the coordinates $(a_1 -1, \cdots, a_n-1)$. Let us now note
that by construction the hyperplane $\CH_{n+1}$ is located at
lattice-distance $\ce*{\frac{d_{n+1} +1}{2}}$ from $O_{n+1}$,
 and also that the face $O_1 O_2 \cdots O_n$ cuts the hyper-parallelogram
 at the following distance from $\tilde O_{n+1}$:
\beq
\left(\sum\limits_{i=1}^n a_i - n\right) - l_n \;=\; \sum\limits_{i=1}^n a_i
- n - \ce*{\frac{\sum\limits_{i=1}^{n+1} a_i - (n+1) }{2}} = \fl*{\frac{d_{n+1} + 1}{2}} \,.
\eeq
Therefore, as in the 2-simplex cases of the previous subsection,
we see that $E_{n+1}= \prod_{i=1}^n a_i - D_{n+1}(-1)$
counts the lattice points in region $I$.

We performed the above computation for the branch $n+1$, which
pictorially corresponds to picking out one particular vertex
$O_{n+1}$ in the simplex $O_1O_2\cdots O_{n+1}$. However, there
is nothing special about $O_{n+1}$ among the $n+1$ vertices, and
we could have done exactly the same for any $O_k$; The final
expression is always the count of the lattice points in
$I=S_n\cap \tilde S_n$. This proves the invariance of $E_k$.

\paragraph{Remarks:}
\begin{itemize}
\item The intersection region $I = S_n \cap \tilde S_n$ always
exists. To see this, we may only check in branch $n+1$ that the
hyperplane $\CH_{n+1}$ is
located closer to $O_{n+1}$ than the face $O_1 \cdots O_n$ is,
{\it i.e.}, the lattice points on $\CH_{n+1}$ have a lower level than those on the face $O_1 \cdots O_n$.
One can easily check this: $$\ce*{\frac{d_{n+1} + 1}{2}} = \ce*{\frac{\sum\limits_{i=1}^{n+1} a_i - 2 a_{n+1} - n +1}{2}}
 \leq \ce*{\frac{\sum\limits_{i=1}^{n+1} a_i - (n+1)}{2}}=  l_n \,.$$
\item We have assumed that the linking numbers $a_i$
obeyed the inequalities~\eqref{firstclass_general}. However,
our argument also works if one of the $n+1$ inequalities is
violated (only one can be violated at a time as in $n=2$ cases).
Suppose that the violation was by $d_{n+1}$, that is, $d_{n+1} < -1$.
Having the length $\ce*{\frac{d_{n+1} + 1 }{2}}$ negative means
that the hyperplane $\CH_{n+1}$ is located outside the hyper-parallelogram.
It is easy to see in this case that $\tilde O_{n+1}$ sits inside
 the simplex $S$. So the intersecting region $I$ is the entire hyper-parallelogram
 and hence, $E_{n+1} = \prod_{i=1}^n a_i$, or equivalently, $D_{n+1}(-1) = 0$.
\end{itemize}

\subsection{Formula for $D_k(-1)={\rm dim}\,i^*_{M_k}(H(X_k))$ }

Before closing the section, let us now enumerate the states explicitly.
As we already have a simple routine for generating Euler numbers,
here we give a similarly simple routine for computing $D_k(-1)$.
Taking the difference between the two
in any one of the branches would then give the total number of Intrinsic
Higgs states. In the next section, we will find yet another method
which does not only enumerate states but also catalog (Intrinsic)
Higgs phase states according to global charges.

Denoting the invariant~\eqref{kindep} of the quiver by $E\equiv E_1=E_2=\cdots=E_{n+1}$,
we can write
\begin{equation}
D_k(-1)= \prod_{i\neq k}a_i- E \ .
\end{equation}
According to  the
first conjecture, these numbers should equal the Coulomb phase indices
up to a sign. While we investigate
the refined version of these quantities in the next section,
we also record here a general counting formula and its very
simple geometric interpretation in terms of the triangular lattice
and hyper-parallelogram introduced in section~5.2.
It turns out that
\beq \label{D1}
D_k(-1) \;=\; \prod\limits_{i\neq k} a_i - V_n(l_n) + \sum\limits_{r=1}^{n+1}
~~(-1)^{r-1}\sum\limits_{i_1 < \cdots < i_r} V_n (l_n- \sum\limits_{\alpha=1}^{r}a_{i_\alpha}) \ ,
\eeq
where $V_n(m)$, for $m\ge 0$, is the number of lattice points
inside the $n$-simplex with edge length $m$, and is defined
to be 1 and 0, respectively, for $m=0$ and $m<0$.

Let us present a sketchy derivation of the above formula. The idea
is first to find a recursive expression for the invariant
quantity $E$, which counts the lattice points in the overlap
region $I$ of the two $n$-simplices, $S_n=O_1 \cdots O_{n+1}$
and $\tilde S_n = \tilde O_1 \cdots \tilde O_{n+1}$, as defined in section~5.2.
For this, one starts from the simplex $S_n$,
in which there are total of $V_n(l_n)$ lattice points.
Amongst them, those lattice points inside the smaller $n$-simplex with
$n+1$ vertices $O_k$, $V_{i,k} (i\neq k)$ must be excluded for each $k$,
while those on its face opposite to $O_k$ 
must still be counted.
After shifting this face by one lattice unit towards $O_k$, the
size of this $k$-th simplex becomes
\begin{equation}
l_n - (a_k- 1) -1 = l_n - a_k \ ,
\end{equation}
and hence, we have to subtract $V_n (l_n - a_k)$ from $V_n(l_n)$, for each $k$.
This, however, is not quite the final answer since a pair from these
$n+1$ simplices, say $i$-th and $j$-th ones, may also overlap
with each other at a smaller simplex with size $l_n - a_i - a_j + 2$
(see, for instance, Figure~\ref{fig:tetrahedron} or~\ref{fig:tetrahedronE}).
Again, one should be careful about the shift of faces for
correct counting, and in this case, it turns out that the
shift by a unit has to be made twice.
Thus, we add $V_n(l_n - a_i - a_j)$ back in, for each pair $(i, j)$
with $l_n - a_i - a_j$ non-negative, etc., and in the end
obtain the expression~\eqref{D1} for $D_k(-1)$.

Note that the discrete volume, $V_n(m)$, can be computed inductively by using the recursion relation
\beq
V_n(m) = V_n(m-1) + V_{n-1}(m) \ ,
\eeq
with the initial condition
\beq
V_1(m)=m+1 \ ,
\eeq
which merely counts the lattice points on a line segment of length $m$.
This leads to the simple expression
\beq\label{V}
V_n(m)= \binom{n+m}{n} \ ,
\eeq
for non-negative $m$ and zero otherwise.
Eq.~\eqref{D1} for $D_k(-1)$ can then be re-written as the alternating sum of
binomial coefficients,
\begin{eqnarray}\label{D2}
D_k(-1) &=& \prod\limits_{i\neq k} a_i -\;\binom{n+l_n}{n} \;+\;
\sum\limits_{i_1} \binom{n+l_n - a_{i_1} }{n}\nonumber\\
&& \;-\;
 \sum\limits_{i_1 < i_2} \binom{n+l_n  - a_{i_1}- a_{i_2} }{n} \; + \; \cdots  \ ,
\end{eqnarray}
where the ellipsis means that we keep adding or subtracting the binomial
coefficients for all collections of subscripts, $i$'s, as long as
$m=l_n  - a_{i_1} - a_{i_2} - \cdots$ is non-negative. One
should take care to truncate this series at places where the
``length" $m$ turns negative, since the binomial function
can give a (unwanted) nonzero number when the upper argument
$n+m$ is a negative integer.

\section{Refined Index $\Omega_{\rm Higgs}^{(k)}(y)$ and the Second Conjecture}

In this section, we finally come to the general proof of the second conjecture
at fully refined level. In the previous section, we already observed that the
degeneracy of the Intrinsic Higgs sector, or equivalently the difference between
the Higgs phase degeneracy and the Coulomb phase degeneracy, is independent of
the branch choice, $k$. The aim here is to elevate that
observation to fully refined level with all angular momentum and $U(1)_R'$
charges manifest. In addition, we will display many examples (in subsection 6.2)
which will illuminate the Higgs phase ground state sector numerically.

Let us recall how the protected spin character reduces
in the Higgs phase to
\begin{eqnarray}
\Omega_{\rm Higgs}^{(k)}(y) = {\rm tr}\;(-1)^{2L_3}y^{2L_3+2{\cal I}}={\rm tr}_{H(M_k)} \;(-1)^{p+q-d_k}y^{2p-d_k}\ .
\end{eqnarray}
Its analog for the Intrinsic Higgs sector,
\begin{eqnarray}
\Omega_{\rm Higgs}^{(k)}(y)\biggr\vert_{\rm Intrinsic}= {\rm tr}_{H(M_k)/ i^*_{M_k}(H(X_k))} \;(-1)^{p+q-d_k}y^{2p-d_k}\ ,
\end{eqnarray}
should be a refined invariant insensitive to wall-crossings.

For cyclic quivers, luckily, these refined indices are also relatively
easy to compute. For this, recall the notion of refined Euler numbers on a
K\"ahler manifold $M$,
\beq \label{6.3}
\chi^p(M) = \sum_{q \geq 0} (-1)^q \; h^{p,q} (M) \,,
\eeq
with the usual Hodge numbers $h^{p,q}={\rm dim}\;H^{p,q}(M)$,
and their generating function\footnote{The usual Euler character is
obtained by substituting $\xi=-1$:
\beq \nn
\chi_\xi (M) \Biggl\vert_{\xi=-1}  = \sum_{p, q \geq 0}
(-1)^{p+q} \;h^{p,q}(M) =  \chi(M) \,.
\eeq}
\beq \label{6.4}
\chi_\xi (M) = \sum_{p \geq 0} \chi^p(M) \;\xi^p \ ,
\eeq 
which we call the refined Euler character.  

Taking $M=M_k$ and comparing $\chi_\xi (M_k)$ against $\Omega_{\rm Higgs}^{(k)}(y)$ 
above, we find the two are related straightforwardly as  
\begin{equation}
\Omega_{\rm Higgs}^{(k)}(y)=(-y)^{-d_k}\chi_{\xi=-y^2}(M_k) \ .
\end{equation}
This identification can be understood from discussions of section 2, 
where we have identified the supercharges that commute with 
$y^{2L_3+2{\cal I}}$ to be $\bar\partial$ and $\bar\partial^\dagger$. 
On the other hand, $\bar\partial+\bar\partial^\dagger$
is nothing but the elliptic operator responsible for the complex
$\bigoplus_q H^{p,q}(M_k)$; the numerical index for this
complex, for each $0\le p\le d_k$, is precisely $\chi^p(M_k)$ in (\ref{6.3}).

While this expression is only an index, it actually carries the
full cohomology information of $M_k$; For any given $p$, only two 
entries in the Hodge diamond contribute to $\chi_\xi(M_k)$, namely $h^{p,p}$ and
$h^{p,d_k-p}$.  The former belongs to $i^*_{M_k}(H(X_k))$ while the 
latter belongs to the Intrinsic Higgs sector; the only exception to 
this occurs for $p=d_k/2$ with $d_k$ even, in which case the two
sector gets mixed in  the single entry $h^{d_k/2,d_k/2}$.
On the other hand,  $i^*_{M_k}(H(X_k))$
is entirely captured by the reduced Poincar\'{e} polynomial, $D_k(x)$,
which together with $\chi_\xi(M_k)$ determines $H^{p,q}(M_k)$ entirely.
In section 6.2, we will present explicit examples of these refined indices
as well as the Hodge diamonds.

Restricting the trace to $i^*_{M_k}(H(X_k))$, we similarly have
\begin{equation}
\Omega_{\rm Higgs}^{(k)}(y)\biggr\vert_{i^*_{M_k}(H(X_k))}=(-y)^{-d_k}
\chi_{\xi=-y^2}(M_k)\biggr\vert_{i^*_{M_k}(H(X_k))}=(-y)^{d_k}D_k(-y)\ ,
\end{equation}
and for the intrinsic sector,
\begin{eqnarray}\label{invfinal}
\Omega_{\rm Higgs}^{(k)}(y)\biggr\vert_{\rm Intrinsic}
&=& \Omega_{\rm Higgs}^{(k)}(y)-  \Omega_{\rm Higgs}^{(k)}(y)
\biggr\vert_{i^*_{M_k}(H(X_k))}\nonumber \\
&=& (-y)^{-d_k}\chi_{\xi=-y^2}(M_k) -(-y)^{-d_k}D_k(-y)\ .
\end{eqnarray}
The second conjecture says that this last expression is an invariant of 
quiver. In subsection~6.3, we will give a complete and rigorous 
proof of this assertion by demonstrating its $k$-independence.

\subsection{Computing the Refined Euler Character }

For cyclic Abelian quivers, we already have a general formula for $D_k(x)$,
so it is now a matter of finding $\chi_\xi(M_k)$.
We shall use the Hirzebruch-Riemann-Roch formula \cite{Hirzebruch} to find that
 $\chi_\xi (M_k)$ have the following integral representation,
\beq \label{chigen}
\chi_\xi (M_k) = \frac{1}{(1+\xi)^n} \int_{X_k} \left[ \prod_{i \neq k}
\left(J_i \frac{1+\xi e^{-J_i}}{1- e^{-J_i}}\right)^{a_i}\right]
\cdot \left(\frac{1-e^{-\sum_{i\neq k} J_i}}{1+\xi e^{-\sum_{i\neq k} J_i}}\right)^{a_k} \,,
\eeq
which is essentially given as the coefficient of
$\prod_{i\neq k} J_i^{a_i-1}$ in the Taylor expansion
of the integrand. As before, $J_i$ are the K\"ahler forms of
$CP^{a_i-1}$ in $X_k$.

To derive the above formula,
let us start from recalling the definitions of
relevant topological quantities and their basic properties.
For a vector bundle $E$ on $X$, we set
\beq
H^{p,q}(X,E) := H^q(X, \Omega^p(E)) \ , ~h^{p,q}(X,E):={\rm dim}\; H^{p,q}(X,E) \ ,
\eeq
and define the holomorphic Euler character
\beq
\chi^p(X,E) := \sum_{q \geq 0} (-1)^q h^{p,q}(X,E) \ ,
\eeq
whose generating function is denoted by
\beq
\chi_\xi (X,E) := \sum_{p \geq 0} \chi^p (X,E) \;\xi^p \,.
\eeq
We write $\chi_\xi(X)$ if $E$ is the trivial holomorphic line bundle, reducing to Eq.~\eqref{6.4}.
Let us also define
\beq
{\rm ch}_\xi (E) := \sum_{p\geq 0} {\rm ch}(\wedge^p E) \;\xi^p \,.
\eeq
Then, for a line bundle $L$ with $c_1(L)=t \in H^2(X, \IZ)$, we have
\bea
{\rm ch}_\xi (L) &=& 1 + \xi e^t \ , \\
{\rm td} (L) &=& \frac {t}{1-e^{-t}} \ ,
\eea
where ${\rm td}(L)$ is the Todd genus of $L$.
We finally note that, if the bundle $E$ is given by the extension sequence
\beq
0 \to E_1 \to E \to E_2 \to 0 \,,
\eeq
then the Todd genus and the Chern character have the following multiplicative properties:
\bea
{\rm td}(E) &=& {\rm td}(E_1) \cdot{\rm td}(E_2) \ , \\
{\rm ch}_\xi(E) &=& {\rm ch}_\xi (E_1) \cdot {\rm ch}_\xi(E_2) \ .
\eea

For a single projective space $X=\IC \IP^{a-1}$, the Euler sequence
\beq
0 \to \CO_X \to \CO_X(1)^{\oplus a} \to \CT X \to 0 \,,
\eeq
leads to
\bea
{\rm td}(\CT X) &=& \left(\frac{J}{1-e^{-J}}\right)^a \ , \\
{\rm ch}_\xi(\CT^*X) &=& \frac{(1+\xi e^{-J})^a}{1+\xi}  \ ,
\eea
where $J$ is the K\"ahler form of $X$.
Due to the multiplicative property of Todd genus and Chern character,
they have a natural generalization to the product of projective
spaces $X_k = \prod_{i\neq k} \IC \IP^{a_i-1}$, which are the ambient varieties of our concern.

We are now ready to address the $\chi_\xi$-character of the
complete intersection $M_k$.
The Adjunction formula dictates that we have the following relations
\bea \label{td}
{\rm td}(\CT M_k) &=& \left[\prod_{i \neq k}
\left(\frac{J_i}{1-e^{J_i}}\right)^{a_i}\right] \cdot
\left(\frac{1-e^{-\sum_{i\neq k} J_i}}{\sum_{i\neq k} J_i}\right)^{a_k} \ , \\
{\rm ch}_\xi (\CT^*M_k) &=& \left[\prod_{i \neq k}
\frac{(1+ \xi e^{-J_i})^{a_i}}{1+\xi} \right] \cdot
\left(\frac{1}{1+\xi e^{-\sum_{i\neq k} J_i}}\right)^{a_k}\ , \label{ch}
\eea
where $J_i$ is the K\"ahler form from each $\IC \IP^{a_i - 1}$ factor in $X_k$.
Therefore, upon applying the Hirzebruch-Riemann-Roch formula~\cite{Hirzebruch}, we have
\bea
\nn \chi_\xi(M_k) &=& \int_{M_k} {\rm td} (\CT M_k) \cdot {\rm ch}_\xi (\CT^* {M_k}) \\
\nn &=& \int_{X_k} {\rm td} (\CT M_k) \cdot {\rm ch}_\xi (\CT^* {M_k})
 \cdot  \Big(\sum_{i\neq k} J_i\Big)^{a_k} \\
&=&  \frac{1}{(1+\xi)^n} \int_{X_k} \left[ \prod_{i \neq k}
\left(J_i \frac{1+\xi e^{-J_i}}{1- e^{-J_i}}\right)^{a_i}\right]
\cdot \left(\frac{1-e^{-\sum_{i\neq k} J_i}}{1+\xi e^{-\sum_{i\neq k} J_i}}\right)^{a_k} \,,
\eea
and hence, arrive at the expression~\eqref{chigen} as promised.
Note that in the second step the Poincar\'e dual of $[M_k]$ was
multiplied to elevate the integral to the ambient space $X_k$,
and in the third step Eqs.~\eqref{td} and~\eqref{ch} have been used.

\subsection{Indices and Hodge Numbers : Numerical Illustrations}

Let us study the refined indices and the resulting Hodge numbers for specific
examples. To understand typical shape of Hodge diamonds, let us first take a
3-gon example with $a_1=a_2=a_3=8$, so that all three branches are identical
with the complex dimension $d=6$. In this case, the Hodge diamond of $h^{p,q}$'s can be drawn as
\begin{equation} \label{888}
\begin{array}{ccccccccccccc}
&&&&&&1&&&&&& \\
&&&&&0&&0&&&&& \\
&&&&0&&2&&0&&&&\\
&&&0&&0&&0&&0&&&\\
&&0&&0&&3&&0&&0&&\\
&0&&0&&0&&0&&0&&0&\\
1\;\;&&322&&6803&&18216&&6803&&322&&\;\;1 \quad .\\
&0&&0&&0&&0&&0&&0&\\
&&0&&0&&3&&0&&0&&\\
&&&0&&0&&0&&0&&&\\
&&&&0&&2&&0&&&&\\
&&&&&0&&0&&&&& \\
&&&&&&1&&&&&& \\
\end{array} \nonumber
\end{equation}
The horizontal line represents (mostly) intrinsic Higgs states, except
four out of the $h^{3,3}=18216=18212+4$ states, that belong to $i^*_M(H(X))$.
This illustrates how the Intrinsic Higgs states and the rest separate
neatly (except at $h^{d/2,d/2}$ when $d$ is even) into middle horizontal and middle vertical
part of the Hodge diamond. This cross-like pattern with horizontal Intrinsic
part and vertical pulled-back part, with possible overlap at the center $h^{d/2,d/2}$
when $d$ is even, is a general feature of cyclic Abelian quivers.

To illustrate what happens when different branches are topologically
distinct, we take  another  3-gon example, with $(a_1, a_2, a_3)=(4,5,6)$,
for which we obtain the following Hodge diamonds in branches 1, 2 and 3,
respectively:
\begin{equation}
\begin{array}{ccccccccccc}
&&&&&1&&&&& \\
&&&&0&&0&&&& \\
&&&0&&2&&0&&&\\
&&0&&0&&0&&0&&\\
&0&&0&&3&&0&&0&\\
0&&0&&26&&26&&0&&0\\
&0&&0&&3&&0&&0&\\
&&0&&0&&0&&0&&\\
&&&0&&2&&0&&&\\
&&&&0&&0&&&& \\
&&&&&1&&&&& \\
\end{array} \ ,
\qquad
\begin{array}{ccccccc}
&&&1&&& \\
&&0&&0&& \\
&0&&2&&0&\\
0&&26&&26&&0\\
&0&&2&&0&\\
&&0&&0&& \\
&&&1&&& \\
\end{array} \ , \qquad
\begin{array}{ccc}
&1& \\
26&&26 \ .\\
&1& \\
\end{array}\nonumber
\end{equation}
Note how the same middle line is repeated. In this case, the
complex dimensions of $M_k$'s are odd, so Coulomb states
do not mix in the middle cohomology, which then represents
the Intrinsic Higgs states entirely. Equivalently, these
data are encoded in the refined index of the Higgs phase.
For example, in branch 1, the index is given by
\beq
\Omega^{(1)}_{\rm Higgs} (y) = - \frac{1}{y^5} - \frac{2}{y^3} +\frac{23}{y} + 23 y - 2y^3 - y^5   \ ,
\eeq
and when restricted to the pulled-back cohomology, by
\beq
\Omega^{(1)}_{\rm Higgs} (y)\biggr\vert_{i^*_{M_1}(H(X_1))} = (-y)^{-d_1}D_1(-y)
=- \frac{1}{y^5} - \frac{2}{y^3} -\frac{3}{y} - 3 y - 2y^3 - y^5   \ .
\eeq
Then the Intrinsic Higgs sector has the following refined index
\begin{eqnarray}
\nn \Omega^{(1)}_{\rm Higgs}(y)\biggr\vert_{\rm Intrinsic}^{(4,5,6)}&=&
\Omega^{(1)}_{\rm Higgs} (y) - \Omega^{(1)}_{\rm Higgs} (y)\biggr\vert_{i^*_{M_1}(H(X_1))} \\
&=& \frac{26}{y} + 26 y \ ,
\end{eqnarray}
which, in this case, describes the invariant middle cohomology in full detail.
Repeating the same exercise for branches 2 and 3, we find
\begin{eqnarray}
\Omega^{(1)}_{\rm Higgs}(y)\biggr\vert_{\rm Intrinsic}^{(4,5,6)}=
\Omega^{(2)}_{\rm Higgs}(y)\biggr\vert_{\rm Intrinsic}^{(4,5,6)}=
\Omega^{(3)}_{\rm Higgs}(y)\biggr\vert_{\rm Intrinsic}^{(4,5,6)}= \frac{26}{y} + 26 y \ ,
\end{eqnarray}
as advertised.
Finally, we record a few more, less-trivial, examples. The first is another 3-gon example
with $(a_1,a_2,a_3)=(15,16,17)$. The total refined Higgs phase
indices in the three branches are given as,
\begin{eqnarray}
\Omega^{(1)}_{\rm Higgs} (y)&=&
1/y^{16}+2/y^{14}+1668/y^{12}+724678/y^{10}+60686568/y^8\nonumber\\
&&+1523273850/y^6+13886938956/y^4+50685934046/y^2 +77668453896\nonumber\\
&&+50685934046y^2+13886938956 y^4+1523273850 y^6\nonumber\\
&&+60686568 y^8+724678 y^{10}+1668 y^{12}+2
y^{14}+y^{16} \ , \nonumber \\
\Omega^{(2)}_{\rm Higgs} (y)&=&1/y^{14}+1667/y^{12}+724677/y^{10}+60686567/y^8 \nonumber\\
&&+1523273849/y^6+13886938955/y^4+50685934045/y^2+77668453895\nonumber\\
&&+50685934045
y^2+13886938955 y^4+1523273849 y^6\nonumber\\
&&+60686567 y^8+724677 y^{10}+1667 y^{12}+y^{14} \ , \nonumber \\
\Omega^{(3)}_{\rm Higgs} (y)&=&1666/y^{12}+724676/y^{10}+60686566/y^8+1523273848/y^6\ ,\nonumber\\
&&+13886938954/y^4+50685934044/y^2+77668453894 \nonumber \\
&&+50685934044y^2+13886938954 y^4+1523273848 y^6\nonumber\\
&&+60686566 y^8+724676 y^{10}+1666 y^{12}\ .
\end{eqnarray}
Similarly, refined Higgs indices for the pulled-back part are,
\begin{eqnarray}
(-y)^{-d_1}D_1(-y)&=&1/y^{16}+2/y^{14}+3/y^{12}+4/y^{10}+5/y^8+6/y^6+7/y^4+8/y^2\nonumber\\
&&+9+8 y^2+7 y^4+6y^6+5 y^8+4 y^{10}+3 y^{12}+2 y^{14}+y^{16} \ ,\nonumber\\
(-y)^{-d_2}D_2(-y)&=&1/y^{14}+2/y^{12}+3/y^{10}+4/y^8+5/y^6+6/y^4+7/y^2\nonumber\\
&&+8+7 y^2+6 y^4+5 y^6+4 y^8+3 y^{10}+2 y^{12}+y^{14}\ ,\nonumber\\
(-y)^{-d_3}D_3(-y)&=&1/y^{12}+2/y^{10}+3/y^8+4/y^6+5/y^4+6/y^2\nonumber \\
&&+7+6 y^2+5 y^4+4 y^6+3 y^8+2 y^{10}+y^{12} \ ,
\end{eqnarray}
from which we find
\begin{eqnarray}
\Omega_{\rm Higgs} (y)\biggl\vert_{\rm Intrinsic}^{(15,16,17)}
&=&\Omega^{(k)}_{\rm Higgs} (y)-(-y)^{-d_k}D_k(-y)\nonumber \\
&=&1665/y^{12}+724674/y^{10}+60686563/y^8+1523273844/y^6\nonumber \\
&&+13886938949/y^4+50685934038/y^2+
77668453887 \nonumber \\
&&+50685934038 y^2+13886938949 y^4+1523273844 y^6\nonumber\\
&&+60686563 y^8+724674 y^{10}+1665 y^{12} \ ,
\end{eqnarray}
independent of $k$, again as advertised. The second example is a 4-gon
case with $(a_1,a_2,a_3,a_4)=(5,6,7,8)$. The total refined Higgs phase
indices in the four branch are give as,
\begin{eqnarray}
\Omega^{(1)}_{\rm Higgs} (y)&=&-1/y^{13}-3/y^{11}
-6/y^9+4415/y^7+362210/y^5+5127653/y^3+18229383/y\nonumber\\
&&+18229383 y+5127653 y^3+362210 y^5+4415 y^7-6 y^9-3 y^{11}-y^{13}\ ,\nonumber \\
\Omega^{(2)}_{\rm Higgs} (y)&=&-1/y^{11}-3/y^9+4419/y^7+362215/y^5+5127659/y^3+18229390/y\nonumber\\
&&+18229390 y+5127659 y^3+362215 y^5+4419 y^7-3 y^9-y^{11}\ ,\nonumber\\
\Omega^{(3)}_{\rm Higgs} (y)&=&-1/y^9+4422/y^7+362219/y^5+5127664/y^3+18229395/y\nonumber\\
&&+18229395 y+5127664 y^3+362219 y^5+4422 y^7-y^9\ , \nonumber\\
\Omega^{(4)}_{\rm Higgs} (y)&=&4424/y^7+362222/y^5+5127668/y^3+18229400/y\nonumber\\
&&+18229400 y+5127668 y^3+362222 y^5+4424 y^7 \ ,
\end{eqnarray}
while refined indices for the pulled-back part are,
\begin{eqnarray}
(-y)^{-d_1}D_1(-y)&=&-1/y^{13}-3/y^{11}-6/y^9-10/y^7-15/y^5-21/y^3-27/y\nonumber\\
&&-27 y-21 y^3-15 y^5-10 y^7-6 y^9-3 y^{11}-y^{13}\ ,\nonumber\\
(-y)^{-d_2}D_2(-y)&=&-1/y^{11}-3/y^9-6/y^7-10/y^5-15/y^3-20/y\nonumber\\
&&-20 y-15 y^3-10 y^5-6 y^7-3 y^9-y^{11}\ ,\nonumber\\
(-y)^{-d_3}D_3(-y)&=&-1/y^9-3/y^7-6/y^5-10/y^3-15/y\nonumber \\
&&-15 y-10 y^3-6 y^5-3 y^7-y^9\ ,\nonumber\\
(-y)^{-d_4}D_4(-y)&=&-1/y^7-3/y^5-6/y^3-10/y\nonumber\\
&&-10 y-6 y^3-3 y^5-y^7\ ,
\end{eqnarray}
which, together, gives
\begin{eqnarray}
\Omega_{\rm Higgs} (y)\biggl\vert_{\rm Intrinsic}^{(5,6,7,8)}
&=&\Omega^{(k)}_{\rm Higgs} (y)-(-y)^{-d_k}D_k(-y)\nonumber \\
&=&4425/y^7+362225/y^5+5127674/y^3+18229410/y\nonumber \\
&&+18229410 y+5127674
y^3+362225 y^5+4425 y^7\ ,
\end{eqnarray}
independent of $k$. The promised invariance can be seen
very clearly.

With consistently large linking numbers, it is also clear that
the degeneracy of the Intrinsic Higgs sector grows very
fast. For example, with linking numbers of mixed sizes, say
$(a_1,a_2,a_3,a_4,a_5)=(2,3,5,8,13)$, we find a relatively
small degeneracy,
\begin{eqnarray}
&&\Omega_{\rm Higgs} (y)\biggl\vert_{\rm Intrinsic}^{(2,3,5,8,13)}=
1261261/y+1261261y\ .
\end{eqnarray}
With more consistently larger numbers, the growth of the
Intrinsic Higgs sector is very rapid. For example, with
the same notation as above, we have
\begin{eqnarray}
&&\Omega_{\rm Higgs} (y)\biggl\vert_{\rm Intrinsic}^{(3,5,7,9,11)}=
\nonumber \\
&&54599524/y^9+3730179061/y^7+63638875882/y^5+379987985704/y^3\nonumber\\
&&+905199873928/y+905199873928y+379987985704y^3\nonumber \\
&&+63638875882y^5+373017906y^7+54599524y^9 \ ,
\end{eqnarray}
and
\begin{eqnarray}
&&\Omega_{\rm Higgs} (y)\biggl\vert_{\rm Intrinsic}^{(5,6,7,8,9)}=
\nonumber \\
&&831775/y^{13}+301581526/y^{11}+22987872352/y^9+575641637000/y^7\nonumber\\
&&+5763858350669/y^5+25595480770735/y^3+53280763215115/y\nonumber\\
&&+53280763215115y+25595480770735y^3+5763858350669y^5\nonumber\\
&&+575641637000y^7+22987872352y^9+301581526y^{11}+831775y^{13} \ ,
\end{eqnarray}
and as a final example,
\begin{eqnarray}
&&\Omega_{\rm Higgs} (y)\biggl\vert_{\rm Intrinsic}^{(8,9,10,11,12)}=
\nonumber \\
&&32294250/y^{22}+58872952926/y^{20}+23086762587054/y^{18}\nonumber\\
&&+3146301650299568/y^{16}+186529800766285403/y^{14}\nonumber\\
&&+5480846262397291070/y^{12}+86780383421802203555/y^{10}\nonumber\\
&&+783408269154731872224/y^8+4192271239441338802849/y^6\nonumber\\
&&+13657486692285216220742/y^4+27560691162972524163666/y^2\nonumber\\
&&+34791235315880411958041+27560691162972524163666y^2\nonumber\\
&&+13657486692285216220742y^4+4192271239441338802849y^6\nonumber\\
&&+783408269154731872224y^8+86780383421802203555y^{10}\nonumber\\
&&+5480846262397291070y^{12}+186529800766285403y^{14}\nonumber\\
&&+3146301650299568y^{16}+23086762587054y^{18}\nonumber\\
&&+58872952926y^{20}+32294250y^{22}\ .
\end{eqnarray}
The growth is expected to be exponential, in general, as is
appropriate for the interpretation of these states as black hole
microstates.

\subsection{Proof of the Second Conjecture}

Now that we understand the general structure of $\Omega_{\rm Higgs}(y)$
and its restriction to the Intrinsic Higgs sector, let us  go ahead
and prove the second conjecture in its full generality. Using the
(already proven) first conjecture, invariance of the refined index
of the Intrinsic Higgs sector amounts to $k$-independence of
\begin{equation}
\Omega_{\rm Higgs}^{(k)}(y)-(-y)^{-d_k}D_k(-y)\quad \leftrightarrow
\quad\Omega_{\rm Higgs}^{(k)}(y)-\Omega_{\rm Coulomb}^{(k)}(y) \ ,
\end{equation}
which can be also stated as the equality condition,
\begin{equation}\label{HHCC}
\Omega_{\rm Higgs}^{(k)}(y)-\Omega_{\rm Higgs}^{(k')}(y)\;\;
=\;\;\Omega_{\rm Coulomb}^{(k)}(y)-\Omega_{\rm Coulomb}^{(k')}(y)
\end{equation}
with arbitrary pairs of branches, $k$ and $k'$, for any given quiver.
In the following we will show that this latter statement holds for
all cyclic Abelian quivers.

On the Higgs side, we start by rewriting (\ref{chigen}) as contour integrals, so
that $\Omega_{\rm Higgs}^{(k)}(y)=(-y)^{-d_k}\chi_{\xi=-y^2}(M_k)$ is equal to
\begin{eqnarray}
  \frac{(-y)^{-d_k}}{(1-y^2)^n} \prod_{i\neq k}\oint_{J_i=0}\frac{dJ_i}{2\pi i}
\left[ \prod_{i \neq k}\left(\frac{1-y^2 e^{-J_i}}{1- e^{-J_i}}\right)^{a_i}\right]
\cdot \left(\frac{1-e^{-\sum_{i\neq k} J_i}}{1-y^2 e^{-\sum_{i\neq k} J_i}}\right)^{a_k}  \ ,
\end{eqnarray}
which maps to, with $\omega_i\equiv e^{-J_i}$,
\begin{eqnarray}\label{O1}
 \frac{(-y)^{-d_k}}{(y^2-1)^n} \prod_{i\neq k}\oint_{\omega_i=1}\frac{d\omega_i}{2\pi i}
\left[ \prod_{i \neq k}
\frac1{\omega_i}\left( \frac{1-y^2 \omega_i}{1- \omega_i}\right)^{a_i}\right]
\cdot \left(\frac{1- \prod_{i \neq k} \omega_i}{1-y^2 \prod_{i \neq k} \omega_i}\right)^{a_k}\ .
\end{eqnarray}
A trick that simplifies the proof enormously is to rewrite the last factor of the integrand
in terms of another contour integral with a dummy variable $\omega_k$ as
\begin{eqnarray}
&&\left(\frac{1- \prod_{i \neq k} \omega_i}{1-y^2 \prod_{i \neq k} \omega_i}\right)^{a_k}\cr\cr\cr
&=&\oint_{\omega_{k}={y^{-2} \prod_{i \neq k} \omega_i^{-1}}} \frac{d\omega_k}{2\pi i}
\left(\frac{1- y^{-2}\omega_k^{-1}}{1-\omega_k^{-1}}\right)^{a_k}\cdot\frac{1}{\omega_k-y^{-2} \prod_{i \neq k} \omega_i^{-1}} \cr\cr\cr
&=&-\frac{\prod_{i\neq k}\omega_i}{y^{2a_k-2}}\oint_{\omega_{k}={y^{-2} \prod_{i \neq k} \omega_i^{-1}}} \frac{d\omega_k}{2\pi i}
\left(\frac{1-y^2\omega_k}{1-\omega_k}\right)^{a_k}\cdot \frac{1}{1-y^2\prod_i\omega_i}
\end{eqnarray}
The integrand has two additional poles at $\omega_k=1$ and $\omega_k=\infty$, so we may
trade this contour integral in favor of two others as,
\begin{eqnarray}
&=&\frac{\prod_{i\neq k}\omega_i}{y^{2a_k-2}}\left(-\oint_{\omega_{k}=\infty}+\oint_{\omega_k=1}\right) \frac{d\omega_k}{2\pi i}
\left(\frac{1-y^2\omega_k}{1-\omega_k}\right)^{a_k}\cdot \frac{1}{1-y^2\prod_i\omega_i}\cr\cr\cr
&=&1\;\;+\;\;\frac{\prod_{i\neq k}\omega_i}{y^{2a_k-2}}\oint_{\omega_k=1} \frac{d\omega_k}{2\pi i}
\left(\frac{1-y^2\omega_k}{1-\omega_k}\right)^{a_k}\cdot \frac{1}{1-y^2\prod_i\omega_i}
\end{eqnarray}
where  the first term ``1"  is from $\omega_k=\infty$ residue. Inserting this
back into (\ref{O1}), we find $\Omega_{\rm Higgs}^{(k)}(y)$ is composed
of two additive pieces. The first piece, from ``1",
\begin{eqnarray}
\frac{(-y)^{-d_k}}{(y^2-1)^n} \prod_{i\neq k}\oint_{\omega_i=1}\frac{d\omega_i}{2\pi i}
\left[ \prod_{i \neq k}
\frac1{\omega_i}\left( \frac{1-y^2 \omega_i}{1- \omega_i}\right)^{a_i}\right]
\end{eqnarray}
depends on the branch choice, $k$, while the second piece
\begin{eqnarray}
&&\frac{(-y)^{-d_k-2a_k+2}}{(y^2-1)^n} \prod_{i}\oint_{\omega_i=1}\frac{d\omega_i}{2\pi i}
\left[ \prod_{i}
\left( \frac{1-y^2 \omega_i}{1- \omega_i}\right)^{a_i}\right]\cdot\frac{1}{1-y^2\prod_i \omega_i} \cr\cr\cr
&=&\frac{(-y)^{n+2-\sum_i a_i }}{(y^2-1)^n} \prod_{i}\oint_{\omega_i=1}\frac{d\omega_i}{2\pi i}
\left[ \prod_{i}
\left( \frac{1-y^2 \omega_i}{1- \omega_i}\right)^{a_i}\right]\cdot\frac{1}{1-y^2\prod_i \omega_i}
\end{eqnarray}
is manifestly independent of $k$.

The first, $k$-dependent, piece
of $\Omega_{\rm Higgs}^{(k)}(y)$ is explicitly integrated with
\begin{eqnarray}
\oint_{\omega_{i}=1}\frac{d\omega_{i}}{2\pi i}
\left( \frac{1-y^2 \omega_i}{1- \omega_i}\right)^{a_k}
\frac{1}{\omega_i}&=&y^{2a_i}-1 \ ,
\end{eqnarray}
and becomes
\begin{eqnarray}
\frac{(-y)^{-d_k}}{(y^2-1)^n}\prod_{i\neq k} (y^{2a_i}-1)
=(-1)^{d_k}y^{a_k}\prod_{i\neq k} \frac{y^{a_i}-y^{-a_i}}{y-y^{-1}} \ .
\end{eqnarray}
Thus, we have the general formula for the refined index in the Higgs phase
for arbitrary cyclic Abelian quiver,
\begin{eqnarray}
\Omega_{\rm Higgs}^{(k)}(y)
&=&(-1)^{d_k}y^{a_k}\prod_{i\neq k} \frac{y^{a_i}-y^{-a_i}}{y-y^{-1}} \\ \nonumber \\
&&+
\frac{(-y)^{n+2-\sum_i a_i }}{(y^2-1)^n} \prod_{i}\oint_{\omega_i=1}\frac{d\omega_i}{2\pi i}
\left[ \prod_{i}
\left( \frac{1-y^2 \omega_i}{1- \omega_i}\right)^{a_i}\right]\cdot\frac{1}{1-y^2\prod_i \omega_i} \ .\nonumber
\end{eqnarray}
As a simple corollary, we have
\begin{eqnarray}\label{O2}
\Omega_{\rm Higgs}^{(k)}(y)-\Omega_{\rm Higgs}^{(k')}(y)
=
(-1)^{d_k-1}\frac{y^{a_k-a_{k'}}-y^{-a_k+a_{k'}}}{y-y^{-1}}
\prod_{i\neq k,k' }\frac{y^{a_i}-y^{-a_i}}{y-y^{-1}}
\ ,
\end{eqnarray}
where we remembered that $(-1)^{d_k}=(-1)^{-n+\sum_i a_i}$ is independent of $k$.
We will presently compare this against the Coulomb phase counterpart.

For the Coulomb phase, we start with Eq.~(\ref{GC}). Without loss
of generality, we may suppose $a\equiv a_k-a_{k'}> 0$: If $a_k<a_{k'}$,
we exchange the two labels, while, for $a_k=a_{k'}$, (\ref{GC}) shows
$\Omega_{\rm Coulomb}^{(k)}(y)-\Omega_{\rm Coulomb}^{(k')}(y)=0$
immediately, so no computation is needed.
Taking the difference between branch $k$ and branch $k'$, we find
\begin{eqnarray}\label{Gky-Gkpy}
&&G_k(y)-G_{k'}(y)
\cr&=&\sum_{\{t_{i\neq k,k'}=\pm1\}}\Big[\prod_{i\neq k,k'}t_i\Big] \cdot
\left[\Theta\Big(\sum_{i\neq k,k'} a_i { t}_i-a\Big)\,
y^{\sum_{i\neq k,k'} a_i { t}_i-a}\right.
\cr&&~~~~~~~~~~~~~~~~~~~~~~~~~~~~~~\left.-
\Theta\Big(\sum_{i\neq k,k'} a_i { t}_i+a\Big)
\,y^{\sum_{i\neq k,k'} a_i { t}_i+a}\right]
\cr&=&\sum_{\{t_{i\neq k,k'}=\pm1\}}\Big[\prod_{i\neq k,k'}t_i\Big] \cdot
\left[\Theta\Big(\sum_{i\neq k,k'} a_i { t}_i- a\Big)\,
y^{\sum_{i\neq k,k'} a_i { t}_i}(y^{-a}-y^{a})\right.
\cr&&~~~~~\left.-
\Big(1-\Theta\Big(\vert \sum_{i\neq k,k'} a_i { t}_i\vert-a\Big)\Big) \, y^{\sum_{i\neq k,k'} a_i { t}_i+a}
-\delta_{\sum_{i\neq k,k'} a_i { t}_i+a,0}\right]\, .
\end{eqnarray}
We perform  a similar trick on $G(y^{-1})$'s and find
\begin{eqnarray}\label{Gkyi-Gkpyi}
&&(-1)^n\left[G_k(y^{-1})-G_{k'}(y^{-1})\right]
\cr&=&(-1)^n\sum_{\{t_{i\neq k,k'}=\pm1\}}\Big[\prod_{i\neq k,k'}t_i\Big] \cdot
\left[\Theta\Big(\sum_{i\neq k,k'} a_i { t}_i- a\Big)\,
y^{-\sum_{i\neq k,k'} a_i { t}_i}(y^{a}-y^{-a})\right.
\cr&&~~~~~\left.-
\Big(1-\Theta\Big(|\sum_{i\neq k,k'} a_i { t}_i|-a\Big)\Big)\, y^{-\sum_{i\neq k,k'} a_i { t}_i-a}
-\delta_{-\sum_{i\neq k,k'} a_i { t}_i-a,0}\right]
\cr&=&-\sum_{\{t_{i\neq k,k'}=\pm1\}}\Big[\prod_{i\neq k,k'}t_i\Big] \cdot
\left[\Theta\Big(-\sum_{i\neq k,k'} a_i { t}_i- a\Big)\,
y^{\sum_{i\neq k,k'} a_i { t}_i}(y^{a}-y^{-a})\right.
\cr&&~~~~~\left.-
\Big(1-\Theta\Big(|\sum_{i\neq k,k'} a_i { t}_i|-a\Big)\Big)\, y^{\sum_{i\neq k,k'} a_i { t}_i-a}
-\delta_{\sum_{i\neq k,k'} a_i { t}_i-a,0}\right]\,,
\end{eqnarray}
where we have taken one more step of flipping the definition, $t_i\rightarrow-t_i$,
at the last two lines.

Combining Eqs.~(\ref{Gky-Gkpy}) and (\ref{Gkyi-Gkpyi}) together, we find a vastly simplified
expression,
\begin{eqnarray}\label{Omega-Omega}
&&\frac{G_k(y)+(-1)^nG_k(y^{-1})-G_{k'}(y)-(-1)^nG_{k'}(y^{-1})}{(y-y^{-1})^n}
\cr&=&(y-y^{-1})^{-n}(y^{-a}-y^{a})\sum_{\{t_{i\neq k,k'}=\pm1\}}
\Big[\prod_{i\neq k,k'}t_i\Big]\, y^{\sum_{i\neq k,k'}a_it_i}
\cr&=&-\frac{y^{a}-y^{-a}}{y-y^{-1}}\prod_{i\neq k,k'}\frac{y^{a_i}-y^{-a_i}}{y-y^{-1}}\,.
\end{eqnarray}
which is already finite at $y=1$ so that the difference of the two
counter polynomials vanishes on its own, $H_{k}(y)-H_{k'}(y)=0$.

The difference of the equivariant Coulomb indices between branches
$k$ and $k'$ is therefore
\begin{eqnarray}\label{OmegaC-OmegaC2}
\Omega_{\rm Coulomb}^{(k)}(y)-\Omega_{\rm Coulomb}^{(k')}(y)
=(-1)^{d_k-1}\frac{y^{a_k-a_{k'}}-y^{-a_k+a_{k'}}}{y-y^{-1}}\prod_{i\neq k,k'}\frac{y^{a_i}-y^{-a_i}}{y-y^{-1}}\ .
\end{eqnarray}
Comparing this against (\ref{O2}), we see the
two expressions are identical. This leaves behind only the case of
$a_k=a_{k'}$, to which the above Coulomb phase procedure does not
extend. However, as noted already, the difference vanishes in this
case, which is mirrored by the Higgs phase result (\ref{O2}) as well.
This establishes (\ref{HHCC}), which in turn guarantees that
$\Omega_{\rm Higgs}^{(k)}(y)-\Omega_{\rm Coulomb}^{(k)}(y)
=\Omega_{\rm Higgs}^{(k)}(y)-(-y)^{-d_k}D_k(-y)$ is an
invariant of the quiver. This generalizes the invariance proof in
section 5 to the refined level.

\section{Conclusion}

In this note, we showed that the two conjectures proposed in Ref.~\cite{Lee:2012sc}
hold for all cyclic Abelian quivers, at a refined level with angular
momentum and $R$-charge information: the Coulomb phase ground states are
in one-to-one correspondence to elements of the pulled-back
ambient cohomology, $i^*_{M_k}(H(X_k))$,
while the remainder of $H(M_k)$, which we call the Intrinsic Higgs sector, is found to
be insensitive to wall-crossing and defines an invariant of the quiver
itself. Along the way, we constructed
the  refined index in the Higgs phase which computes the protected spin
character of BPS states in four dimensions, and offered a routine for
determining the entire Hodge diamond of the Higgs phase vacuum moduli
space. Also found is a simple arithmetic formula for the Coulomb
phase degeneracy.

As we already mentioned in section 2,
all known BPS states, constructed to date from
four-dimensional $N=2$ field theories, are $SU(2)_R$ singlets.
Partly based on such observations, it has been speculated that BPS
states are neutral under $R$ symmetry and thus are entirely
classified by its angular momentum representations \cite{Gaiotto:2010be}.
As far as we know, on the other hand, quivers that construct field
theory BPS states admit no Intrinsic Higgs sector; it may as well be
that $SU(2)_R$ singlet property is the hallmark of the Coulomb phase states.
The latter view is consistent with our finding $i^*_{M_k}(H(X_k))
=\bigoplus_p i^*_{M_k}(H^{p,p}(X_k))$ in all of our examples.
For the Intrinsic Higgs states, however, vanishing $R$-charges
would be strange, if not logically impossible; no classifying global
charge would remain since they are  inherently angular momentum
singlets \cite{Lee:2012sc,Bena:2012hf}.\footnote{The vanishing angular momentum
as a criteria for single-center black hole states, as opposed to multi-center ones,
was first proposed and tested extensively for 1/4 BPS black hole microstates
in Refs.~\cite{Sen:2009vz,Sen:2010mz}.} Indeed,
among several things we learned in this note is that, for the Intrinsic
Higgs states, $U(1)_R'$-charge are typically nontrivial, which ultimately
means that these states are classified by $SU(2)_R$ of the underlying
four-dimensional $N=2$ theory.

This suggests a perhaps more physical, if less precise, criterion to
separate the Coulomb states and the Intrinsic Higgs states (modulo
those singlet states in $H^{d/2,d/2}$ for even $d$). The former states are
all in the representation of type
\begin{equation}
[{\rm1/2\; hyper}]\otimes (J,0)\ ,
\end{equation}
while the latter states are in
\begin{equation}
[{\rm1/2\; hyper}]\otimes (0, I)\ ,
\end{equation}
for some collection of $J$'s and $I$'s. We expect that $SU(2)_R$
representations are encoded entirely in ${\cal I}$ eigenvalues
and degeneracies thereof. Assuming single-center black hole
interpretation of the latter, in particular, this implies that
the microstates of BPS black holes of four-dimensional $N=2$
theories are classified by $SU(2)_R$ multiplets. Finding explicit
constructions of the corresponding black hole microstates
from string theory models and comparing the resulting $R$-charge
content against Eq.~(\ref{invfinal}) should be most illuminating.

Although we have found  an explicit and simple
characterization that distinguish between Coulomb-like
states and Intrinsic Higgs states, it remains a little
mysterious from the Higgs phase viewpoint why the former
states suffer wall-crossing while the latters do not.
Physically, the single-center black hole interpretation
for the Intrinsic Higgs states, as opposed to multi-center
one, goes a long way explaining their invariance but
then again, it is not exactly transparent why such
a dichotomy of BPS states occurs and also why the
field theory BPS states seemingly belong only to
the former class of states.

At the level of quiver quantum mechanics, the disparity between
the two phases can be understood from how ground state dynamics
relate to the full quiver dynamics. For large absolute values of
FI parameters, the Higgs phases are quite reliable as the vacuum
manifold tends to be large and the truncated massive directions
are very massive. For the Coulomb phase, things are a little more
subtle, however. Small absolute values of FI constants favor this
phase and wall-crossing physics become physically more transparent
here, yet the naive truncation to the conventional Coulombic vacuum
moduli space is dynamically unjustified due to small mass gaps
along classically massive directions \cite{Kim:2011sc}. For index computation,
this problem can be evaded via an index preserving deformation
\cite{Kim:2011sc}, but things become qualitatively more difficult
for quivers that accept the so-called scaling solutions
\cite{Denef:2007vg,Manschot:2011xc}.
The usual adoption of flat kinetic term is no longer justified,
even for the purpose of index computations, near the origin
where two or more charge centers approach each other arbitrarily close.
There, the topology can be quite different than naively assumed \cite{Lee:2011ph},
casting some doubt on the usual prescription. Such subtleties
may explain why the Coulomb phase fails to capture the entire
low energy aspects of quiver dynamics.

On a more mathematical side, we can also ask how this relates
to the proposal of Kontsevich and Soibelman, who offered
a simple algebraic structure that is supposed to
capture the wall-crossing behavior in a universal manner
\cite{KS,Gaiotto:2008cd,Gaiotto:2009hg}.
In this approach, indices on two sides of a given marginal
stability wall enters, respectively, a string of operator
product as exponents. Wall-crossing data is recovered, then,
by demanding that two such operator products equal each
other, which constrains exponents of one side given those
on other side. The equivalence of this proposal with physically
derived wall-crossing formulae has been tested extensively,
but only for examples where the Intrinsic Higgs sectors
are absent.\footnote{See Ref.~\cite{Sen:2011aa} for the most
complete comparison, to date.}
This leads to the question of whether and how the information of the Intrinsic
Higgs sector enters this algebraic formulation.
While the underlying multi-center physics of wall-crossing
is very clearly encoded in the Coulomb phase, the universal nature of the
algebra suggests that the total index rather than just
the Coulomb phase index would enter the algebra.
Whether this is true or not needs to be confirmed, to
begin with. Assuming an affirmative answer, it would
also raise a question of whether and how
Kontsevich-Soibelman algebra might know about the quiver
invariants in some natural manner.

Apart from the  next obvious task of verifying these
conjectures for  general quivers,
also of some interest is generalization of this story
to 1/4 BPS states of four-dimensional $N=4$ theories \cite{Bergman:1997yw}.
In fact, the multi-centered nature and the intuitive
understanding of wall-crossing of what we now call
Coulomb phase states was first discovered in the context of
such states \cite{Lee:1998nv,Bak:1999da,Bak:1999ip},
and quiver representations are also available for them, albeit
with complications from having adjoint Higgs fields.
Analogs of the refined index for these 1/4 BPS states
have been identified recently \cite{Sen:2012hv}, which
may be explored along the same line as here.

\vskip 1cm
\centerline{\bf\large Acknowledgments}
\vskip 5mm\noindent
We are grateful to Ashoke Sen for useful comments and  indebted
to HwanChul Yoo for helpful advice on combinatorics issues.
This work is supported by the National Research foundation of Korea (NRF)
funded by the Ministry of Education, Science and Technology
with the grant number 2010-0013526 and also in part (PY) via
the Center for Quantum Spacetime with grant number 2005-0049409.

\appendix


\begin{thebibliography}{99}



\bibitem{Denef:2002ru}
  F.~Denef,
  ``{Quantum Quivers and Hall/Hole Halos},''
  JHEP {\bf 0210} (2002) 023
  [arXiv:hep-th/0206072].

\bibitem{Prasad:1975kr}
  M.K.~Prasad and C.M.~Sommerfield,
``{An Exact Classical Solution for the 't Hooft Monopole andthe
Julia-Zee
  Dyon},''
  Phys. Rev. Lett.  {\bf 35} (1975) 760;
  E.B.~Bogomolny,
  ``{Stability of Classical Solutions},''
  Sov. J. Nucl. Phys.  {\bf 24} (1976) 449
  [Yad.\ Fiz.\  {\bf 24} (1976) 861].

\bibitem{Seiberg:1994rs}
  N.~Seiberg and E.~Witten,
``{Monopole Condensation, And Confinement In N=2
Supersymmetric Yang-Mills
  Theory},''
  Nucl. Phys.  B {\bf 426} (1994) 19
  [Erratum-ibid.\  B {\bf 430} (1994) 485]
  [arXiv:hep-th/9407087];
  N.~Seiberg and E.~Witten,
``{Monopoles, Duality and Chiral Symmetry Breaking in N=2
Supersymmetric
  QCD},''
  Nucl. Phys. B {\bf 431} (1994) 484
  [arXiv:hep-th/9408099].

\bibitem{Ferrari:1996sv}
  F.~Ferrari and A.~Bilal,
``{The Strong-Coupling Spectrum of the Seiberg-Witten
Theory},''
  Nucl. Phys.  B {\bf 469}, 387 (1996)
  [arXiv:hep-th/9602082].


\bibitem{Lee:1998nv}
  K.M.~Lee and P.~Yi,
``{Dyons in N=4 Supersymmetric Theories and Three Pronged
Strings},''
  Phys. Rev.  {\bf D58 } (1998)  066005.
  [hep-th/9804174].

\bibitem{Bak:1999da}
  D.~Bak, C.~K.~Lee, K.~M.~Lee and P.~Yi,
  ``Low-energy dynamics for 1/4 BPS dyons,''
  Phys.\ Rev.\ D {\bf 61} (2000) 025001
  [hep-th/9906119].

  \bibitem{Gauntlett:1999vc}
  J.P.~Gauntlett, N.~Kim, J.~Park and P. Yi
``{Monopole Dynamics and BPS Dyons N=2 Super Yang-Mills
Theories},''
  Phys. Rev. {\bf D61 } (2000)  125012.
  [hep-th/9912082].

  \bibitem{Denef:2000nb}
  F.~Denef,
  ``{Supergravity Flows and D-brane Stability},''
  JHEP {\bf 0008} (2000) 050
  [arXiv:hep-th/0005049].



\bibitem{Ritz:2000xa}
  A.~Ritz, M.~A.~Shifman, A.~I.~Vainshtein and M.~B.~Voloshin,
  ``Marginal stability and the metamorphosis of BPS states,''
  Phys.\ Rev.\ D {\bf 63} (2001) 065018
  [hep-th/0006028].

\bibitem{Argyres:2001pv}
  P.~C.~Argyres and K.~Narayan,
  ``String webs from field theory,''
  JHEP {\bf 0103} (2001) 047
  [hep-th/0101114].


\bibitem{Sen:2011aa}
  A.~Sen,
  ``Equivalence of Three Wall Crossing Formulae,''
  arXiv:1112.2515 [hep-th].


\bibitem{Denef:2007vg}
  F.~Denef and G.W.~Moore,
  ``{Split States, Entropy enigmas, Holes and Halos},''
  [arXiv:hep-th/0702146].

\bibitem{Kim:2011sc}
  H.~Kim, J.~Park, Z.~Wang and P.~Yi,
  ``Ab Initio Wall-Crossing,''  JHEP {\bf 1109}, 079 (2011)
  [arXiv:1107.0723 [hep-th]].

\bibitem{Lee:2011ph}
  S.~Lee and P.~Yi,
  ``Framed BPS States, Moduli Dynamics, and Wall-Crossing,''
  JHEP {\bf 1104} (2011) 098
  [arXiv:1102.1729 [hep-th]].




\bibitem{Manschot:2011xc}
  J.~Manschot, B.~Pioline and A.~Sen,
  ``A Fixed point formula for the index of multi-centered N=2 black holes,''
  JHEP {\bf 1105}, 057 (2011)
  [arXiv:1103.1887 [hep-th]].

\bibitem{Lee:2012sc}
  S.~-J.~Lee, Z.~-L.~Wang and P.~Yi,
  ``Quiver Invariants from Intrinsic Higgs States,''
  arXiv:1205.6511 [hep-th].


\bibitem{Bena:2012hf}
  I.~Bena, M.~Berkooz, J.~de Boer, S.~El-Showk and D.~V.~d.~Bleeken,
  ``Scaling BPS Solutions and pure-Higgs States,''
  arXiv:1205.5023 [hep-th].




\bibitem{Gaiotto:2010be}
  D.~Gaiotto, G.~W.~Moore and A.~Neitzke,
  ``Framed BPS States,''
  arXiv:1006.0146 [hep-th].





\bibitem{Stern:2000ie}
  M.~Stern and P.~Yi,
  ``{Counting Yang-Mills Dyons with Index Theorems},''
  Phys. Rev. {\bf D62 } (2000)  125006.
  [hep-th/0005275].


\bibitem{deBoer:2008zn}
  J.~de Boer, S.~El-Showk, I.~Messamah and D.~Van den Bleeken,
  ``{Quantizing N=2 Multicenter Solutions},''
  JHEP {\bf 0905} (2009) 002
  [arXiv:0807.4556 [hep-th]].






\bibitem{Manschot:2010qz}
  J.~Manschot, B.~Pioline and A.~Sen,
  ``Wall Crossing from Boltzmann Black Hole Halos,''
  [arXiv:1011.1258 [hep-th]].

  \bibitem{Weinberg:2006rq}
  E.~J.~Weinberg and P.~Yi,
  ``Magnetic Monopole Dynamics, Supersymmetry, and Duality,''
  Phys.\ Rept.\  {\bf 438}, 65 (2007)
  [arXiv:hep-th/0609055].

\bibitem{GH}
P.~Griffiths and J.~Harris, ``Principles of Algebraic Geometry,'' John Wiley \& Sons, 1978.




\bibitem{Hirzebruch}
F.~Hirzebruch,
``Topological Methods in Algebraic Geometry,'' Springer, 1966.



\bibitem{Sen:2009vz}
  A.~Sen,
  ``Arithmetic of Quantum Entropy Function,''
  JHEP {\bf 0908} (2009) 068
  [arXiv:0903.1477 [hep-th]].

\bibitem{Sen:2010mz}
  A.~Sen,
  ``How Do Black Holes Predict the Sign of the Fourier Coefficients of Siegel Modular Forms?,''
  Gen.\ Rel.\ Grav.\  {\bf 43} (2011) 2171
  [arXiv:1008.4209 [hep-th]].

\bibitem{KS}
  M. Kontsevich and Y. Soibelman,
``{Stability Structures, Motivic Donaldson-Thomas Invariants
and Cluster Transformations},'' [arXiv:0811.2435]


\bibitem{Gaiotto:2008cd}
  D.~Gaiotto, G.W.~Moore and A.~Neitzke,
``{Four-dimensional Wall-crossing via Three-dimensional
Field Theory},''
  Commun. Math. Phys. {\bf 299} (2010) 163
  [arXiv:0807.4723 [hep-th]].

\bibitem{Gaiotto:2009hg}
  D.~Gaiotto, G.~W.~Moore and A.~Neitzke,
  ``Wall-crossing, Hitchin Systems, and the WKB Approximation,''
  arXiv:0907.3987 [hep-th].

\bibitem{Bergman:1997yw}
  O.~Bergman,
  ``Three pronged strings and 1/4 BPS states in N=4 superYang-Mills theory,''
  Nucl.\ Phys.\ B {\bf 525} (1998) 104
  [hep-th/9712211].



\bibitem{Bak:1999ip}
  D.~Bak, K.~-M.~Lee and P.~Yi,
  ``Quantum 1/4 BPS dyons,''
  Phys.\ Rev.\ D {\bf 61} (2000) 045003
  [hep-th/9907090].

\bibitem{Sen:2012hv}
  A.~Sen,
  ``BPS Spectrum, Indices and Wall Crossing in N=4 Supersymmetric Yang-Mills Theories,''
  arXiv:1203.4889 [hep-th].



\end{thebibliography}
\end{document}